\documentclass[aps,prl,twocolumn,superscriptaddress,showpacs,floatfix,10pt]{revtex4-2}

\usepackage{graphicx}
\usepackage{ifthen}
\usepackage{xcolor}
\usepackage{soul}
\usepackage{bm}
\usepackage{braket}
\usepackage{multirow}
\usepackage{hyperref}
\hypersetup{
 pdfnewwindow=true, colorlinks=true,
 linkcolor=blue, anchorcolor=blue,
 citecolor=blue, filecolor=blue,
 menucolor=blue, urlcolor=blue}

\def\ve{\varepsilon}
\def\tc{T_{\rm c}}
\def\ef{\ve_{\rm F}}
\def\efeff{E_{\rm F}^{\rm eff}}

\def\o{\omega}
\def\op{\omega^\prime}

\def\l{\lambda}
\def\lV{\lambda^{V}}
\newcommand{\wlg}{\ensuremath{\omega_{\log}}}
\def\AL{A_{\lambda}}
\def\RV{R_{V}}
\newcommand{\aaff}{\ensuremath{\alpha^2F(\omega)}}
\newcommand{\aaffv}{\ensuremath{\alpha^2F^{V}(\o,\op)}}

\newcommand{\runin}[1]{\smallskip\noindent\textit{#1---}\ }

\begin{document}

\title{Anharmonicity and Nonadiabaticity in Hydride Superconductors}

\author{Shashi B. Mishra}
\email{mshashi125@gmail.com}
\affiliation{Department of Physics, Binghamton University-SUNY, Binghamton, NY, 13902, USA}
\thanks{Current address: Department of Electrical and Computer Engineering, University of Maryland, College Park, Maryland 20742, USA}
\author{Francesco Belli}
\affiliation{Department of Chemistry, State University of New York at Buffalo, Buffalo, NY, 14260-3000, USA}
\author{Eva Zurek}
\affiliation{Department of Chemistry, State University of New York at Buffalo, Buffalo, NY, 14260-3000, USA}
\author{Elena R. Margine}
\email{rmargine@binghamton.edu}
\affiliation{Department of Physics, Binghamton University-SUNY, Binghamton, NY, 13902, USA}
\date{\today}
\keywords{Hydrides, anharmonicity, electron-phonon coupling, vertex corrections, nonadiabaticity, superconductivity}

\begin{abstract}
We study superconductivity in representative hydrides using anharmonic phonons, electron-phonon vertex corrections, and full-bandwidth Eliashberg theory. The high-pressure binary hydrides H$_3$S, YH$_6$, and YH$_9$ must be treated with both anharmonic and nonadiabatic corrections, whereas the ambient-pressure PdH/PdD/PdT series is strongly anharmonic but remains adiabatic, reproducing the inverse isotope effect without sizable vertex contributions. LaBeH$_8$ exhibits weak anharmonicity, while vertex corrections reduce the critical temperature ($\tc$) by approximately 4~K, leaving the predicted $\tc$ above experiment. To identify when treatments beyond harmonic, adiabatic Migdal-Eliashberg theory are required, we introduce the anharmonic renormalization $\AL$ and the vertex ratio $\RV$ as material-specific diagnostics. 
\end{abstract}

\maketitle

\runin{Introduction}Hydrogen-rich compounds under pressure are the most prominent example of superconductors with high critical temperatures, $\tc$s, whose discovery and development have been guided by theory~\cite{Ashcroft2004,Duan2014,Li2015,Liu2017,Peng2017,Boeri2022,Zurek2019, Zurek:2021k,Cataldo2021}. This includes $\tc$ values above 200~K in H$_3$S~\cite{Drozdov2015} and the clathrate phases YH$_6$ and YH$_9$~\cite{Troyan2021,Kong2021,Wang2022}, and a near-room-temperature $\tc$ in LaH$_{10}$~\cite{Drozdov2019,Somayazulu2019}. Most recently, superconductivity was predicted~\cite{Zhang2022} and experimentally observed~\cite{Song2023} in the submegabar ternary hydride LaBeH$_8$, with $\tc\approx110$~K at 80~GPa. These developments were all anticipated or rationalized by Migdal-Eliashberg (ME) calculations~\cite{Migdal1958,Eliashberg1960,Allen1983}. Yet, quantitative agreement is not uniform across this family of superconductors: in several benchmarks, the calculated transition temperatures of hydrides remain above experiment even in state-of-the-art treatments~\cite{Pellegrini2024}.

Several refinements of ME theory have been proposed as solutions to these discrepancies, but their behavior has usually been examined separately. Quantum anharmonic lattice dynamics is essential when light hydrogen atoms undergo large zero-point motion, strongly renormalizing phonon spectra and electron-phonon (e-ph) coupling~\cite{Ganguly1976,Griessen1982,Jena1984,Monacelli2021}. In H$_3$S, for example, anharmonicity hardens the hydrogen-derived modes and substantially lowers both the e-ph coupling and $\tc$~\cite{Errea2015,Mishra2026}. Full-bandwidth (FBW) Eliashberg theory provides a complementary refinement by retaining the energy dependence of the electronic density of states (DOS), which cannot be replaced by its Fermi-level, $\ef$, value when sharp features such as van Hove singularities (vHs) lie nearby~\cite{Sano2016,Lucrezi2024}. Improved methods to estimate the Coulomb pseudopotential, either an \textit{ab initio} evaluation or its renormalization to a Matsubara frequency cutoff in the Eliashberg solution, offer another route to quantitative accuracy~\cite{Pellegrini2024,Kogler2025,Ding2025}.

Beyond anharmonic lattice dynamics, another limitation of conventional ME theory is the neglect of e-ph vertex corrections. Migdal's theorem justifies this approximation when phonon energies are small compared with the electronic energy scale relevant to pairing~\cite{Migdal1958}.
This separation is not guaranteed in hydrides, where hydrogen-derived phonons can reach energies of 100--250~meV, becoming comparable to the energy scale over which the electronic structure varies near $\ef$. H$_3$S is the prototypical case, with its sharp van Hove features producing a rapidly varying DOS near $\ef$~\cite{Sano2016,Gorkov2018,Akashi2020,Talantsev2022}, such that first-order vertex corrections affect the pairing interaction~\cite{Mishra2025,Mishra2026,Durajski2016non-adia}. Anharmonicity and nonadiabaticity therefore test different assumptions underlying ME theory: the former primarily requires going beyond the harmonic Born-Oppenheimer description by renormalizing the phonons~\cite{Monacelli2021} and the Eliashberg function \aaff{}, whereas the latter directly challenges Migdal's approximation~\cite{Migdal1958} by modifying the e-ph pairing kernel~\cite{Botti2002,Schrodi2020,Mishra2025}. This distinction motivates a material-specific question: which approximation limits the accuracy of ME predictions in a given hydride?

In this Letter, we calculate anharmonic phonons within the Stochastic Self-Consistent Harmonic Approximation using bespoke machine learning interatomic potentials (ML/SSCHA)~\cite{Belli2025}. We combine these phonons with first-order e-ph vertex corrections in full-bandwidth Eliashberg theory, as implemented in EPW~\cite{Mishra2025}, to determine from first principles the superconducting gap and $\tc$ of representative hydrides. Application of this framework to H$_3$S, YH$_6$, YH$_9$, LaBeH$_8$, and PdH/PdD/PdT distinguishes hydrides requiring both anharmonic and nonadiabatic corrections from an anharmonic but adiabatic limit and a weak-anharmonicity regime in which a finite vertex correction remains insufficient. We quantify these regimes using two dimensionless descriptors, the anharmonic renormalization $\AL$ and the vertex ratio $\RV$, which classify each hydride according to the corrections required.

\runin{Methods}We performed density-functional theory and density-functional perturbation theory calculations with Quantum~{\small ESPRESSO}~\cite{Giannozzi2017,Baroni2001} using optimized norm-conserving Vanderbilt pseudopotentials~\cite{Hamann2013} and the Perdew-Burke-Ernzerhof exchange-correlation functional~\cite{Perdew1996}. Anharmonic phonons were obtained from ML/SSCHA calculations following the procedure outlined in Ref.~\cite{Belli2025}. The e-ph matrix elements were interpolated in the Wannier representation~\cite{Giustino2007,Marzari2012,Pizzi2020,Marazzo2024}, and the superconducting properties were computed with EPW~\cite{Margine2013,Lee2023}. The overall computational workflow is summarized in Fig.~S1~\cite{SI}. The isotropic Eliashberg equations were solved in the FBW formulation~\cite{Lucrezi2024,Lee2023,Mishra2025} for all compounds. For all high-pressure hydrides, we chose the conventional value $\mu^{*}=0.13$, commonly adopted for such systems~\cite{Liu2017,Peng2017,Liu2025,Cataldo2021,Zurek2019}, whereas the lower value $\mu^{*}=0.085$ was applied to the ambient-pressure Pd hydrides, following earlier studies~\cite{Rowe1986,Griessen1982,Errea2013}. All calculations were performed for structures relaxed at the static-lattice pressure $P$; the corresponding quantum-corrected pressure $\tilde{P}$, which adds the SSCHA quantum-nuclear contribution to the stress of the same cell, is reported alongside it in Fig.~\ref{fig:framework}(a) and Table~S1~\cite{SI}.
Additional computational details are provided in the Supplemental Material (SM)~\cite{SI}. 


\begin{figure}[!t]
\centering
\includegraphics[width=\linewidth]{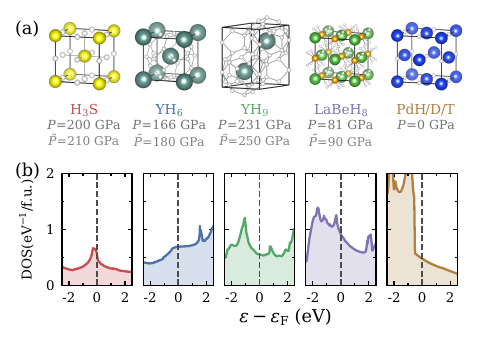}
\caption{(a) Crystal structures of the hydrides studied here, labeled with pressure ($P$) and the quantum-corrected SSCHA pressure ($\tilde{P}$). Small white spheres denote H; large spheres denote S (yellow), Y (dark green), La (light green), Be (orange), and Pd (blue). (b) Electronic density of states, with $\ef$ indicated by the dashed line.} 
\label{fig:framework}
\end{figure}

\runin{Materials landscape}The phases studied here [Fig.~\ref{fig:framework}(a); Table~S1~\cite{SI}] were chosen to span distinct bonding environments: a three-dimensional covalent S--H network in H$_3$S~\cite{Duan2014,Drozdov2015}; weakly bonded, negatively charged hydrogen networks stabilized by electropositive Y atoms in YH$_6$ and YH$_9$~\cite{Li2015,Troyan2021,Kong2021,Wang2022}; Be--H molecular units embedded in a positively charged La sublattice in LaBeH$_8$~\cite{Zhang2022,Song2023}; and octahedrally coordinated H atoms occupying interstitial sites in the face-centered cubic Pd lattice in PdH~\cite{Worsham1957,McLennan2008}. These bonding motifs lead to qualitatively different electronic structures at $\ef$ [Fig.~\ref{fig:framework}(b); Figs.~S2 and S3; Table~S2~\cite{SI}]. 

In H$_3$S, a pronounced DOS peak arising from strongly hybridized H-$s$/S-$p$ bands lies just below $\ef$~\cite{Akashi2020}. This rapid variation near $\ef$ makes H$_3$S sensitive to nonadiabatic e-ph vertex corrections~\cite{Mishra2025,Mishra2026} and requires a FBW Eliashberg treatment that includes states away from $\ef$~\cite{Sano2016,Lucrezi2024,Mishra2025}. In the clathrate-like Y hydrides, the metallic states near $\ef$ have mixed H-$s$/Y-$d$ character. YH$_6$ exhibits a DOS that varies smoothly around $\ef$~\cite{Li2015,Lucrezi2024}. YH$_9$ has a greater H-derived character but a slightly lower DOS at $\ef$, which lies in a valley between nearby peaks. However, the shorter H--H contacts are accompanied by larger hopping amplitudes than in YH$_6$ (Table~S2~\cite{SI}), indicating enhanced electronic coupling within the hydrogen framework. This stronger coupling helps offset the lower DOS and contributes to the higher $\tc$ of YH$_9$. LaBeH$_8$ and PdH both exhibit a comparatively gradual, approximately linear variation of the DOS near $\ef$. In LaBeH$_8$, $\ef$ lies on the descending edge of a broad shoulder~\cite{Wang2024}, while in PdH it lies on the smoothly decreasing tail of a broad free-electron-like $s$ band, above the main Pd-$4d$/H-$1s$ peak located about 1~eV lower in energy~\cite{Papaconstantopoulos1978,Gupta1978,Houari2014}.

The variation of each DOS profile near $\ef$ defines the effective electronic energy scale $\efeff$ relevant to nonadiabaticity, as described in the SM and tabulated in Table~S3~\cite{SI}.
The susceptibility to nonadiabatic vertex corrections is governed by the ratio of the characteristic phonon energy to this electronic scale. In the high-pressure hydrides, the large phonon energies and, in some cases, a small $\efeff$ associated with rapid DOS variations can make this ratio appreciable. In the ambient-pressure Pd hydrides, the much smaller phonon energies keep the ratio small and suppress vertex corrections. By contrast, anharmonic renormalization depends on how strongly the potential-energy surface departs from a quadratic form and on the extent of nuclear quantum motion.

\begingroup
\squeezetable
\begin{table}[t]
\centering
\caption{Electron-phonon couplings, descriptors, and critical temperatures across the hydride family. The listed $\tc$ values are isotropic FBW results obtained with harmonic (har), anharmonic (anh), and anharmonic$+$vertex (anh$+$ver) treatments, using $\mu^{*}=0.13$ for the high-pressure hydrides and $0.085$ for the ambient-pressure Pd hydrides. Static-lattice and quantum-corrected pressures are listed in Fig.~\ref{fig:framework}(a) and Table~S1~\cite{SI}. Experimental $\tc$ values are from Ref.~\cite{Drozdov2015} for H$_3$S at 200~GPa, Ref.~\cite{Troyan2021} for YH$_6$ at 166~GPa, Ref.~\cite{Kong2021} for YH$_9$ at 231~GPa, Ref.~\cite{Song2023} for LaBeH$_8$ at 80~GPa, and Refs.~\cite{Skoskiewicz1972,Stritzker1972,Rowe1986,Schirber1984} for PdH/PdD/PdT at ambient pressure.} 
\label{tab:summary}
\begin{ruledtabular}
\begin{tabular}{l cccc cccc}
 & \multicolumn{4}{c}{coupling and descriptors} & \multicolumn{4}{c}{$T_c$ (K)} \\
\cline{2-5}\cline{6-9}
Material & $\lambda_{\rm har}$ & $\lambda_{\rm anh}$ & $A_\lambda$ & $R_V$ & har & anh & anh$+$ver & expt. \\
\colrule
H$_3$S & 2.29 & 1.49 & 0.35 & 0.32 & 244 & 196 & 184 & 184 \\
YH$_6$ & 2.36 & 1.74 & 0.26 & 0.14 & 260 & 233 & 226 & 224 \\
YH$_9$ & 3.09 & 1.83 & 0.41 & 0.14 & 265 & 244 & 231 & 239 \\
LaBeH$_8$ & 1.66 & 1.62 & 0.02 & 0.11 & 164 & 163 & 159 & 110 \\
PdH & 5.56 & 0.48 & 0.91 & 0.08 & -- & 8.0 & 8.0 & 8--9 \\
PdD & 5.20 & 0.57 & 0.89 & 0.08 & -- & 10.5 & 10.5 & 10--11 \\
PdT & 5.11 & 0.62 & 0.88 & 0.08 & -- & 11 & 11 & 12 \\
\end{tabular}
\end{ruledtabular}
\end{table}
\endgroup

\runin{Anharmonic phonon renormalization}Harmonic and anharmonic phonon dispersions and phonon DOSs are compared in Fig.~S4~\cite{SI}, and the corresponding e-ph couplings are summarized in Table~\ref{tab:summary}. For H$_3$S, the ML/SSCHA spectrum agrees closely with the ZG special-displacement result used in our earlier work (Fig.~S5~\cite{SI}). The H-derived branches reach 200--250~meV in the high-pressure hydrides but stay below approximately 60~meV in the Pd series. In H$_3$S, YH$_6$, and YH$_9$, quantum anharmonicity hardens the H-derived optical branches and suppresses the e-ph coupling by approximately 25--40\%, from $\l_{\rm har}=2.29$, 2.36, and 3.09 to $\l_{\rm anh}=1.49$, 1.74, and 1.83, respectively. The YH$_6$ value agrees closely with an earlier SSCHA~\cite{Troyan2021} and stochastic path-integral results~\cite{Ding2025}, both reporting $\l_{\rm anh}=1.71$ at 165~GPa (Table~S4~\cite{SI}). Such anharmonic hardening and suppression of the e-ph coupling are characteristic of hydrides with symmetric H-bonding environments~\cite{belli2026chemical}.

Part of the anharmonic renormalization of $\l$ comes from the pressure shift induced by quantum-nuclear effects~\cite{Monacelli2021,belli2022impact}. For dense hydrides, these quantum-pressure corrections amount to approximately 10--20~GPa over the 100--300~GPa pressure range. Because the e-ph coupling is highly sensitive to the equilibrium volume, accounting for this shift is a relevant source of uncertainty in $\l_{\rm anh}$ and $\tc$. In YH$_6$ and YH$_9$, the impact of this correction is particularly pronounced because $\tc$ exhibits a strong pressure dependence~\cite{Troyan2021,Kong2021,Du2023,Ding2025}. Specifically, the pressure renormalizations induced by quantum nuclear effects for the systems studied here amount to an extra 10~GPa for H$_3$S, 14~GPa for YH$_6$, 19~GPa for YH$_9$, and 9~GPa for LaBeH$_8$. These corrections are included in the $\tilde{P}$ values reported in Fig.~\ref{fig:framework} and Table~S1~\cite{SI}. 

In the Pd hydrides, anharmonicity removes the harmonic instabilities and markedly hardens the H-derived optical rattling modes, with the largest shift in PdH and a progressively weaker renormalization for the heavier isotopes~\cite{Errea2013,Worsham1957,Bianco2026} (Fig.~S4~\cite{SI}). 
The resulting phonon spectra are validated against Raman and inelastic neutron scattering measurements~\cite{Sherman1977,Rowe1974,Rowe1986}. The anharmonic calculations accurately reproduce the acoustic branches and the overall optical-mode dispersion. The calculated longitudinal-optical maxima are only 5--10\% higher in energy than those measured in the substoichiometric crystals~\cite{Rowe1974,Rowe1986}.
The corresponding anharmonic e-ph couplings increase with isotope mass, with $\l_{\rm anh}=0.48$, 0.57, and 0.62 for PdH, PdD, and PdT, respectively, reflecting progressively weaker phonon hardening in the heavier isotopes. The pronounced anharmonic renormalization of the phonon spectra is consistent with the highly symmetric bonding environment~\cite{belli2026chemical}.

\begin{figure}[!t]
    \centering
    \includegraphics[width=\linewidth]{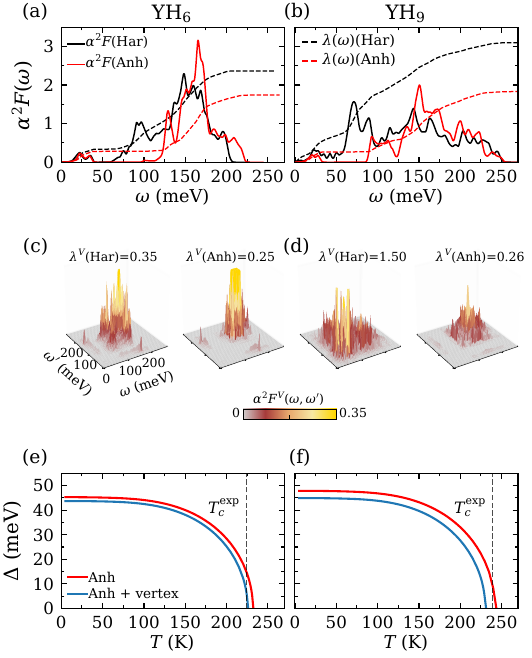}
    \caption{Eliashberg and superconducting properties of YH$_6$ (left) and YH$_9$ (right). (a,b) Eliashberg spectral function, \aaff{}, (solid lines) and cumulative e-ph coupling $\l(\o)$ (dashed lines) for harmonic (black) and anharmonic (red) phonons. (c,d)~Corresponding vertex spectral functions, \aaffv{} and integrated nonadiabatic couplings $\lV$. (e,f) Isotropic superconducting gap $\Delta(T)$ obtained from adiabatic and nonadiabatic calculations with anharmonic phonons, with the experimental $\tc$ values indicated by dashed vertical lines.}
    \label{fig:spectral}
\end{figure}

LaBeH$_8$ lies in a distinct regime of weak anharmonicity. Our anharmonic calculations give only a slight reduction in the e-ph coupling, from $\l_{\rm har}=1.66$ to $\l_{\rm anh}=1.62$, while an earlier study reported an increase from 1.43 to 1.59 at 100~GPa~\cite{Dong2025}. Despite the opposite trends, both studies predict a substantially weaker anharmonic renormalization than in H$_3$S and the Y and Pd hydrides. This is because the system, even if very similar to LaBH$_8$~\cite{belli2022impact}, lies in the transition zone between symmetric and asymmetric bonding environments, where quantum effects harden the phonons and partially compensate for the softening associated with structural changes~\cite{belli2026chemical}.

\runin{Electron-phonon vertex corrections}The e-ph vertex correction captures deviations from Migdal's approximation underlying conventional ME theory~\cite{Migdal1958,Allen1983}. The leading-order vertex diagram contains two phonon and three electron propagators~\cite{Botti2002,Schrodi2020}. Its effect on pairing is encoded in the two-frequency vertex spectral function \aaffv{}, whose double integral yields the nonadiabatic coupling $\lV=4\int d\o \int d\op \alpha^2F^{V}(\o,\op)/(\o\op)$~\cite{Mishra2025}. The magnitude of the vertex correction is governed by the adiabaticity ratio $\eta=\wlg/\efeff$, where $\efeff$ is the characteristic electronic energy scale relevant to pairing. In H$_3$S, the van Hove feature sets $\efeff$ at approximately 40--50~meV~\cite{Mishra2026,Gorkov2018}, well below $\omega_{\rm log}^{\rm anh}=127$~meV (Table~S3~\cite{SI}).

\runin{Y hydrides: strong anharmonicity and moderate vertex corrections}
Figure~\ref{fig:spectral} compares the conventional and vertex spectral functions and their corresponding e-ph couplings in the yttrium clathrates. In both compounds, anharmonicity redistributes the conventional \aaff{} toward higher frequencies, with a stronger overall suppression in YH$_9$. The vertex spectra exhibit a more pronounced contrast between the two materials. In YH$_6$, the near-diagonal distribution is largely preserved, and $\lV$ decreases moderately from $\lV_{\rm har}=0.35$ to $\lV_{\rm anh}=0.25$. In YH$_9$, the broad harmonic spectrum becomes much more localized, bringing $\lV_{\rm har}=1.50$ down to $\lV_{\rm anh}=0.26$. Omitting this renormalization would therefore overestimate $\lV$ in YH$_9$ by nearly a factor of six. The three-dimensional \aaffv{} surfaces for H$_3$S and LaBeH$_8$ are provided in Fig.~S6~\cite{SI}. 

\begin{figure}[!t]
    \centering
    \includegraphics[width=\linewidth]{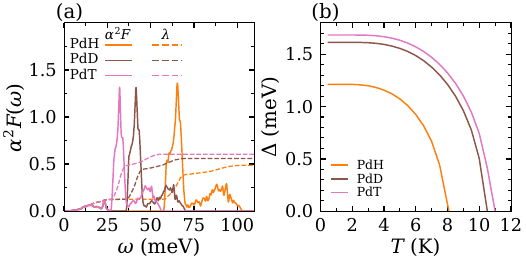}
    \caption{(a)~Eliashberg spectral function \aaff{} (solid) and cumulative $\l(\o)$ (dashed), and (b) superconducting gap $\Delta(T)$ for anharmonic phonons in PdH, PdD, and PdT.} 
    \label{fig:pd}
\end{figure}

The combined impact of anharmonicity and vertex corrections on the superconducting properties is shown in Fig.~\ref{fig:spectral}(e,f). At the conventional value $\mu^{*}=0.13$, anharmonicity lowers $\tc$ from 260 to 233~K in YH$_6$ at 166~GPa and from 265 to 244~K in YH$_9$ at 231~GPa, while first-order vertex corrections further reduce these values to 226 and 231~K, respectively. These results are in close agreement with the measured values of 224~K for YH$_6$ at 166~GPa~\cite{Troyan2021} and 239~K for YH$_9$ at 231~GPa~\cite{Kong2021} (Table~\ref{tab:summary}). The YH$_6$ result is particularly notable because previous harmonic and anharmonic calculations did not reproduce the measured $\tc$~\cite{Troyan2021,Ding2025,Lucrezi2024,Kogler2025}, as discussed in the SM and summarized in Table~S4~\cite{SI}. For YH$_6$, the residual overestimate after anharmonic renormalization is nearly eliminated by including nonadiabatic vertex corrections, consistent with our earlier results for H$_3$S and D$_3$S~\cite{Mishra2025,Mishra2026}. Adopting $\mu^{*}=0.11$ as in Ref.~\cite{Ding2025} increases the vertex-corrected $\tc$ of YH$_9$ to 240~K, essentially matching experiment~\cite{Kong2021}, but also raises that of YH$_6$ to 235~K, exceeding the measured 224~K~\cite{Troyan2021}. The isotropic gap functions $\Delta(T)$ for H$_3$S, YH$_6$, and YH$_9$ using harmonic and anharmonic phonons, with and without first-order vertex corrections, are shown in Fig.~S7~\cite{SI}.

\runin{LaBeH$_8$: weak anharmonicity and a modest vertex correction} LaBeH$_8$ occupies a physically distinct regime. The first-order e-ph vertex correction remains modest compared with those of the other high-pressure hydrides considered here, lowering the anharmonic FBW $\tc$ by 4~K, from 163 to 159~K (Fig.~S7~\cite{SI}). This value remains above the measured $\tc\approx110$~K at 80~GPa~\cite{Song2023}. Previous isotropic Eliashberg calculations at 80 and 100~GPa using harmonic or anharmonic phonons likewise overestimate experiment~\cite{Wang2024,Dong2025,Kogler2025}, including calculations with three Coulomb treatments based on \emph{ab initio} derived $\mu^{*}$ and $\mu$ parameters or an explicit isotropic Coulomb interaction~\cite{Kogler2025} (Table~S4~\cite{SI}). Nonlinear e-ph corrections~\cite{Bianco2026} and anisotropic ME treatments~\cite{Margine2013,Mishra2024,Tomassetti2024} are not obvious remedies because they may further increase $\tc$. Future work should, for instance, assess whether higher-order e-ph vertex corrections can provide the additional suppression needed to bring the calculated $\tc$ into agreement with experiment.

\runin{Inverse isotope effect in the Pd hydrides}
In the Pd hydrides, the inverse isotope effect originates from the mass dependence of anharmonic phonon renormalization. Because anharmonic hardening is greatest in PdH, it suppresses the e-ph coupling most strongly for the lightest isotope~\cite{Ganguly1976,Papaconstantopoulos1978,Griessen1982,Yussouff1995}. Consequently, $\l$ and $\tc$ both increase with isotope mass (Table~\ref{tab:summary} and Fig.~\ref{fig:pd}), reversing the conventional trend expected from the mass scaling of harmonic phonon frequencies~\cite{Rowe1986,Errea2013}. Our anharmonic FBW calculations with $\mu^{*}=0.085$ yield $\tc=8.0$, 10.5, and 11~K for PdH, PdD, and PdT, respectively. These results are in excellent agreement with the reported experimental ranges of 8--9 and 10--11~K for PdH and PdD, respectively~\cite{Skoskiewicz1972,Stritzker1972,Schirber1974,Hemmes1989,Miller1975,Rowe1986}. The agreement with experiment contrasts with previous anharmonic SSCHA calculations, which underestimated $\tc$ at the same $\mu^{*}=0.085$~\cite{Errea2013}. The larger e-ph couplings found here are associated primarily with enhanced \aaff{} weight from the softer lowest H-derived optical branch. A detailed comparison with previous linear and nonlinear e-ph calculations, along with sensitivity of our calculated $\tc$ values to $\mu^{*}$, is provided in the SM~\cite{SI}. The low phonon energies keep the adiabaticity ratio $\eta$ small, resulting in a weak first-order vertex correction, with $\lV_{\rm anh}\simeq0.04$--$0.05$ (Fig.~S8~\cite{SI}) and $\RV\simeq0.08$, that lowers $\tc$ by less than 0.05~K (Fig.~S9~\cite{SI}). The Pd hydrides therefore lie in an anharmonic yet adiabatic regime.

\begin{figure}[!t]
    \centering
    \includegraphics[width=0.95\linewidth]{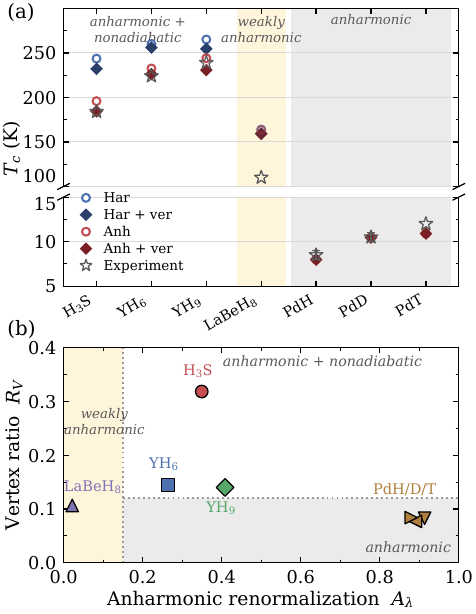}
    \caption{(a) Superconducting $\tc$ obtained from isotropic FBW Eliashberg theory using harmonic (blue) and anharmonic (red) phonons, without (open circles) and with (filled diamonds) first-order vertex corrections, compared with experiment (stars). The corresponding numerical values are listed in Table~\ref{tab:summary}. (b) Classification of the hydrides according to the anharmonic renormalization $\AL$ and the vertex ratio $\RV$; the dotted lines are guides to the eye delineating the different regimes. }
    \label{fig:hierarchy}
\end{figure}

\runin{$\tc$ hierarchy and material descriptors} 
Figure~\ref{fig:hierarchy}(a) summarizes how anharmonicity and first-order vertex corrections modify $\tc$ across the hydrides studied. The dependence of the calculated $\tc$ values on $\mu^{*}$ is shown in Fig.~S11~\cite{SI}. The results fall into three distinct regimes. In H$_3$S, YH$_6$, and YH$_9$, anharmonicity substantially reduces the $\tc$ overestimate from harmonic ME theory, while the vertex correction provides an additional reduction that brings the predictions closer to experiment. For H$_3$S at 200~GPa, $\tc$ decreases from 244~K to 196~K upon including anharmonicity and to 184~K after including the vertex correction, in excellent agreement with the measured value of 184~K at 200~GPa~\cite{Drozdov2015}. LaBeH$_8$ remains substantially above experiment because anharmonicity changes $\tc$ only weakly and the vertex correction lowers it by 4~K, leaving the combined corrections insufficient to resolve the discrepancy. In the Pd hydrides, by contrast, anharmonic ME theory alone reproduces the inverse isotope trend, while the vertex contribution remains small. 

These three regimes reflect the breakdown of two distinct approximations underlying conventional ME theory: the harmonic treatment of the lattice, probed by anharmonicity, and the neglect of e-ph vertex corrections, probed by nonadiabaticity. We quantify these effects using the anharmonic renormalization $\AL=|\l_{\rm anh}-\l_{\rm har}|/\l_{\rm har}$ and the vertex ratio $\RV=|\lV_{\rm anh}|/\l_{\rm anh}$ (Table~\ref{tab:summary} and Table~S3~\cite{SI}). The former measures the relative change in the e-ph coupling induced by anharmonic lattice dynamics, whereas the latter quantifies the first-order vertex contribution relative to $\l_{\rm anh}$. The adiabaticity ratio $\eta=\wlg/\efeff$ provides the corresponding energy-scale criterion for the importance of vertex corrections (Fig.~S10~\cite{SI}). The two descriptors separate the same three regimes into distinct regions of the $\AL$--$\RV$ plane [Fig.~\ref{fig:hierarchy}(b)]: the Pd hydrides have the largest $\AL\approx0.9$ but the smallest $\RV\approx0.08$; H$_3$S combines an intermediate $\AL=0.35$ with the largest vertex ratio, $\RV=0.32$; YH$_6$ and YH$_9$ have comparable $\RV\approx0.14$ but different anharmonic renormalizations, $\AL=0.26$ and 0.41, respectively; and LaBeH$_8$ combines negligible anharmonicity, $\AL=0.02$, with a finite vertex correction, $\RV=0.11$.

\runin{Classification and outlook} We establish a first-principles framework for identifying the material-dependent hierarchy of corrections needed for quantitative predictions of $\tc$. Across the hydrides studied, this approach reproduces the observed $\tc$ trends and yields close agreement with experiment, with LaBeH$_8$ remaining the notable exception. The newly introduced descriptors $\AL$ and $\RV$ quantify the relative magnitudes of anharmonic and first-order vertex corrections and provide material-specific diagnostics that complement the adiabaticity ratio $\eta$ and recently introduced measures of anharmonicity~\cite{belli2026chemical}. These diagnostics are broadly applicable to phonon-mediated superconductors and provide a practical guide for identifying when the harmonic and adiabatic approximations break down.

\runin{Acknowledgments}S.M. and E.R.M. acknowledge the National Science Foundation (NSF) under Award No. OAC-251383. E.Z.\ and F.B.\ acknowledge funding from the U.S.\ Department of Energy, Office of Science, Fusion Energy Sciences Award No.\ DE-SC0020340 and the NSF Award No.\ DMR-2136038. The authors acknowledge the Texas Advanced Computing Center (TACC) at The University of Texas at Austin (http://www.tacc.utexas.edu) and the Center for Computational Research at SUNY Buffalo (http://hdl.handle.net/10477/79221) for providing computational resources that have contributed to the research results reported within this paper.

\runin{Data availability} The first-principles input and output data supporting the findings of this work are openly available in a Figshare repository~\cite{figshare_link}.

\bibliography{pap}

@misc{SI,
note = {{See the Supplemental Material for the computational workflow and details, electronic and vibrational properties including a real-space Wannier analysis, vertex spectral functions and material descriptors, superconducting gaps and $\mu^{*}$ sensitivity, and additional analysis.}}
}

@misc{figshare_link,
  author       = {Mishra, Shashi B. and Belli, Francesco and Zurek, Eva and Margine, Elena R.},
  title        = {{Dataset: Anharmonicity and Nonadiabaticity in Hydride Superconductors}},
  year         = {2026},
  publisher    = {Figshare},
  note         = {Dataset (temporary DOI, to be finalized upon publication)},
  doi          = {10.6084/m9.figshare.31136635.v1},
  url          = {https://doi.org/10.6084/m9.figshare.31136635.v1}
}

@article{Du2023,
  title = {{Superconducting phases of ${\mathrm{YH}}_{9}$ under pressure}},
  author = {Du, Mingyang and Li, Zonglun and Duan, Defang and Cui, Tian},
  journal = {Phys. Rev. B},
  volume = {108},
  issue = {17},
  pages = {174507},
  numpages = {7},
  year = {2023},
  month = {Nov},
  publisher = {American Physical Society},
  doi = {10.1103/PhysRevB.108.174507},
  url = {https://link.aps.org/doi/10.1103/PhysRevB.108.174507}
}

@article{Hemmes1989,
  title = {{Isotope effects and pressure dependence of the ${T}_{c}$ of superconducting stoichiometric PdH and PdD synthesized and measured in a diamond anvil cell}},
  author = {Hemmes, H. and Driessen, A. and Griessen, R. and Gupta, M.},
  journal = {Phys. Rev. B},
  volume = {39},
  issue = {7},
  pages = {4110--4118},
  numpages = {0},
  year = {1989},
  month = {Mar},
  publisher = {American Physical Society},
  doi = {10.1103/PhysRevB.39.4110},
  url = {https://link.aps.org/doi/10.1103/PhysRevB.39.4110}
}

@article{Sherman1977,
title = {Raman studies of hydrogen vibrational modes in palladium},
journal = {Phys. Lett. A},
volume = {62},
number = {5},
pages = {353-355},
year = {1977},
issn = {0375-9601},
doi = {https://doi.org/10.1016/0375-9601(77)90439-X},
url = {https://www.sciencedirect.com/science/article/pii/037596017790439X},
author = {R. Sherman and H.K. Birnbaum and J.A. Holy and M.V. Klein},
}

@Article{Tomassetti2024,
author ={Tomassetti, Charlsey R. and Kafle, Gyanu P. and Marcial, Edan T. and Margine, Elena R. and Kolmogorov, Aleksey N.},
title  ={Prospect of high-temperature superconductivity in layered metal borocarbides},
journal ={J. Mater. Chem. C},
year  ={2024},
volume  ={12},
issue  ={13},
pages  ={4870-4884},
publisher  ="The Royal Society of Chemistry",
doi  ={10.1039/D4TC00210E},
url  ={http://dx.doi.org/10.1039/D4TC00210E}
}

@article{belli2022impact,
  title = {{Impact of ionic quantum fluctuations on the thermodynamic stability and superconductivity of ${\mathrm{LaBH}}_{8}$}},
  author = {Belli, Francesco and Errea, Ion},
  journal = {Phys. Rev. B},
  volume = {106},
  issue = {13},
  pages = {134509},
  numpages = {10},
  year = {2022},
  month = {Oct},
  publisher = {American Physical Society},
  doi = {10.1103/PhysRevB.106.134509},
  url = {https://link.aps.org/doi/10.1103/PhysRevB.106.134509}
}

@article{belli2026chemical,
  title={A chemical bonding based descriptor for predicting the role of anharmonicity induced by quantum nuclear effects in hydride superconductors},
  author={Belli, Francesco and Zurek, Eva and Errea, Ion},
  journal={npj Comput. Mater.},
  year={2026},
  volume={12},
  pages={100},
  doi={10.1038/s41524-026-01973-7},
  publisher={Nature Publishing Group UK London},
  url={https://doi.org/10.1038/s41524-026-01973-7}
}

@Article{Zurek:2021k,
  author = {Hilleke, K. P. and Zurek, E.},
  title  = {Tuning Chemical Precompression: Theoretical Design and Crystal Chemistry of Novel Hydrides in the Quest for Warm and Light Superconductivity at Ambient Pressures},
  year   = {2022},
  journal = {J. Appl. Phys.},
  volume = {131},
  pages = {070901 (1-19)},
  url    = {https://doi.org/10.1063/5.0077748},
}

@article{Kong2021,
  title={{Superconductivity up to 243 K in the yttrium-hydrogen system under high pressure}},
  author={Kong, Panpan and Minkov, Vasily S and Kuzovnikov, Mikhail A and Drozdov, Alexander P and Besedin, Stanislav P and Mozaffari, Shirin and Balicas, Luis and Balakirev, Fedor Fedorovich and Prakapenka, Vitali B and Chariton, Stella and others},
  journal={Nat. comm.},
  volume={12},
  number={1},
  pages={5075},
  year={2021},
  publisher={Nature Publishing Group UK London},
  doi={https://doi.org/10.1038/s41467-021-25372-2}
}

@article{Mishra2026,
author = {Mishra, Shashi B. and Margine, Elena R.},
title = {Nonadiabatic and Anharmonic Effects in High-Pressure H3S and D3S Superconductors},
journal = {Annalen der Physik},
volume = {538},
number = {1},
pages = {e00553},
doi = {https://doi.org/10.1002/andp.202500553},
url = {https://onlinelibrary.wiley.com/doi/abs/10.1002/andp.202500553},
year = {2026}
}

@article{Errea2013,
  title = {First-Principles Theory of Anharmonicity and the Inverse Isotope Effect in Superconducting Palladium-Hydride Compounds},
  author = {Errea, Ion and Calandra, Matteo and Mauri, Francesco},
  journal = {Phys. Rev. Lett.},
  volume = {111},
  issue = {17},
  pages = {177002},
  numpages = {5},
  year = {2013},
  month = {Oct},
  publisher = {American Physical Society},
  doi = {10.1103/PhysRevLett.111.177002},
  url = {https://link.aps.org/doi/10.1103/PhysRevLett.111.177002}
}

@article{Troyan2021,
author = {Troyan, Ivan A. and Semenok, Dmitrii V. and Kvashnin, Alexander G. and Sadakov, Andrey V. and Sobolevskiy, Oleg A. and Pudalov, Vladimir M. and Ivanova, Anna G. and Prakapenka, Vitali B. and Greenberg, Eran and Gavriliuk, Alexander G. and Lyubutin, Igor S. and Struzhkin, Viktor V. and Bergara, Aitor and Errea, Ion and Bianco, Raffaello and Calandra, Matteo and Mauri, Francesco and Monacelli, Lorenzo and Akashi, Ryosuke and Oganov, Artem R.},
title = {{Anomalous High-Temperature Superconductivity in YH$_6$}},
journal = {Adv. Mater.},
volume = {33},
number = {15},
pages = {2006832},
doi = {https://doi.org/10.1002/adma.202006832},
url = {https://advanced.onlinelibrary.wiley.com/doi/abs/10.1002/adma.202006832},
year = {2021}
}

@article{Migdal1958,
  title={Interaction between electrons and lattice vibrations in a normal metal},
  author={Migdal, A. B.},
  journal={Sov. Phys. JETP},
  volume={7},
  pages={996-1001},
  year={1958},
  url={http://jetp.ras.ru/cgi-bin/dn/e_007_06_0996.pdf}
}

@article{Eliashberg1960,
  title = {Interactions between Electrons and Lattice Vibrations in a Superconductor},
  author={Eliashberg, GM},
  journal = {Sov. Phys. JETP},
  volume = {11},
  issue = {3},
  pages = {696--702},
  year = {1960},
  url={https://www.w2agz.com/Library/Classic%20Papers%20in%20Superconductivity/Eliashberg,%20e-p%20Interactions%20in%20SCs,%20Sov-Phys%20JETP%2011,%20696%20(1960).pdf}
}

@article{Duan2014,
  title={{Pressure-induced metallization of dense (H$_2$S)$_2$H$_2$ with high-T$_{\rm c}$ superconductivity}},
  author={Duan, Defang and Liu, Yunxian and Tian, Fubo and Li, Da and Huang, Xiaoli and Zhao, Zhonglong and Yu, Hongyu and Liu, Bingbing and Tian, Wenjing and Cui, Tian},
  journal={Sci. Rep.},
  volume={4},
  number={1},
  pages={6968},
  year={2014},
  url = {https://doi.org/10.1038/srep06968},
  publisher={Nature Publishing Group UK London}
}

@article{Schrodi2020,
  title={{Full-bandwidth Eliashberg theory of superconductivity beyond Migdal's approximation}},
  author={Schrodi, Fabian and Oppeneer, Peter M and Aperis, Alex},
  journal={Phys. Rev. B},
  volume={102},
  number={2},
  pages={024503},
  year={2020},
  publisher={APS},
  doi = {10.1103/PhysRevB.102.024503},
  url={https://doi.org/10.1103/PhysRevB.102.024503}
}

@article{Mishra2024,
  title = {{Stability-superconductivity map for compressed Na-intercalated graphite}},
  author = {Mishra, Shashi B. and Marcial, Edan T. and Debata, Suryakanti and Kolmogorov, Aleksey N. and Margine, Elena R.},
  journal = {Phys. Rev. B},
  volume = {110},
  issue = {17},
  pages = {174508},
  numpages = {13},
  year = {2024},
  month = {Nov},
  publisher = {American Physical Society},
  doi = {10.1103/PhysRevB.110.174508},
  url = {https://link.aps.org/doi/10.1103/PhysRevB.110.174508}
}

@article{Giustino2017,
  title = {Electron-phonon interactions from first principles},
  author = {Giustino, Feliciano},
  journal = {Rev. Mod. Phys.},
  volume = {89},
  issue = {1},
  pages = {015003},
  numpages = {63},
  year = {2017},
  month = {Feb},
  publisher = {American Physical Society},
  doi = {10.1103/RevModPhys.89.015003},
  url = {https://link.aps.org/doi/10.1103/RevModPhys.89.015003}
}

@article{Pellegrini2024,
  title={Ab initio methods for superconductivity},
  author={Pellegrini, Camilla and Sanna, Antonio},
  journal={Nat. Rev. Phys.},
  volume={6},
  number={8},
  pages={509--523},
  year={2024},
  publisher={Nature Publishing Group UK London},
  url={https://www.nature.com/articles/s42254-024-00738-9},
  doi={10.1038/s42254-024-00738-9}
}

@article{Bianco2026,
  url = {https://arxiv.org/abs/2603.03492v1},
  author={Bianco, Raffaello and Errea, Ion},
  title = {Enhanced superconductivity in palladium hydrides by non-perturbative electron-phonon effects}, 
  publisher = {arXiv},
  journal = {arXiv:2603.03492},
  year = {2026}
}

@article{Errea2015,
  title = {High-Pressure Hydrogen Sulfide from First Principles: A Strongly Anharmonic Phonon-Mediated Superconductor},
  author = {Errea, Ion and Calandra, Matteo and Pickard, Chris J. and Nelson, Joseph and Needs, Richard J. and Li, Yinwei and Liu, Hanyu and Zhang, Yunwei and Ma, Yanming and Mauri, Francesco},
  journal = {Phys. Rev. Lett.},
  volume = {114},
  issue = {15},
  pages = {157004},
  numpages = {5},
  year = {2015},
  month = {Apr},
  publisher = {American Physical Society},
  doi = {10.1103/PhysRevLett.114.157004},
  url = {https://link.aps.org/doi/10.1103/PhysRevLett.114.157004}
}

@article{Cataldo2021,
  title = {{$\mathrm{La}{\mathrm{BH}}_{8}$: Towards high-${T}_{c}$ low-pressure superconductivity in ternary superhydrides}},
  author = {Di Cataldo, Simone and Heil, Christoph and von der Linden, Wolfgang and Boeri, Lilia},
  journal = {Phys. Rev. B},
  volume = {104},
  issue = {2},
  pages = {L020511},
  numpages = {6},
  year = {2021},
  month = {Jul},
  publisher = {American Physical Society},
  doi = {10.1103/PhysRevB.104.L020511},
  url = {https://link.aps.org/doi/10.1103/PhysRevB.104.L020511}
}

@article{Song2023,
  title = {{Stoichiometric Ternary Superhydride ${\mathrm{LaBeH}}_{8}$ as a New Template for High-Temperature Superconductivity at 110 K under 80 GPa}},
  author = {Song, Yinggang and Bi, Jingkai and Nakamoto, Yuki and Shimizu, Katsuya and Liu, Hanyu and Zou, Bo and Liu, Guangtao and Wang, Hongbo and Ma, Yanming},
  journal = {Phys. Rev. Lett.},
  volume = {130},
  issue = {26},
  pages = {266001},
  numpages = {8},
  year = {2023},
  month = {Jun},
  publisher = {American Physical Society},
  doi = {10.1103/PhysRevLett.130.266001},
  url = {https://link.aps.org/doi/10.1103/PhysRevLett.130.266001}
}

@article{Botti2002,
  title = {Nonadiabatic theory of the superconducting state},
  author = {Botti, M. and Cappelluti, E. and Grimaldi, C. and Pietronero, L.},
  journal = {Phys. Rev. B},
  volume = {66},
  issue = {5},
  pages = {054532},
  numpages = {10},
  year = {2002},
  month = {Aug},
  publisher = {American Physical Society},
  doi = {10.1103/PhysRevB.66.054532},
  url = {https://link.aps.org/doi/10.1103/PhysRevB.66.054532}
}

@article{Durajski2016non-adia,
  title={{Quantitative analysis of nonadiabatic effects in dense H$_3$S and PH$_3$ superconductors}},
  author={Durajski, Artur P},
  journal={Sci. Rep.},
  volume={6},
  number={1},
  pages={38570},
  year={2016},
  publisher={Nature Publishing Group UK London},
  url={https://doi.org/10.1038/srep38570},
  doi={10.1038/srep38570}
}

@article{Akashi2020,
  title = {{Archetypical ``push the band critical point'' mechanism for peaking of the density of states in three-dimensional crystals: Theory and case study of cubic ${\mathrm{H}}_{3}\mathrm{S}$}},
  author = {Akashi, Ryosuke},
  journal = {Phys. Rev. B},
  volume = {101},
  issue = {7},
  pages = {075126},
  numpages = {18},
  year = {2020},
  month = {Feb},
  publisher = {American Physical Society},
  doi = {10.1103/PhysRevB.101.075126},
  url = {https://link.aps.org/doi/10.1103/PhysRevB.101.075126}
}

@article{Lee2023,
  title = {Electron–phonon physics from first principles using the {EPW} code},
  author = {Lee, Hyungjun and Poncé, Samuel and Bushick, Kyle and  Hajinazar, Samad and Lafuente-Bartolome, Jon and Leveillee, Joshua and Lian, Chao and Lihm, Jae-Mo and Macheda, Francesco and Mori, Hitoshi and Paudyal, Hari and Sio, Weng Hong and Tiwari, Sabyasachi and Zacharias, Marios and Zhang, Xiao and Bonini, Nicola and Kioupakis, Emmanouil and Margine, Elena R. and Giustino, Feliciano},
  journal = {npj Comput. Mater.},
  volume = {9},
  issue = {1},
  pages = {2057-3960},
  year = {2023},
  url = {https://doi.org/10.1038/s41524-023-01107-3}
}

@article{Kogler2025,
title = {{IsoME: Streamlining high-precision Eliashberg calculations}},
journal = {Comp. Phys. Comm.},
volume = {315},
pages = {109720},
year = {2025},
issn = {0010-4655},
doi = {https://doi.org/10.1016/j.cpc.2025.109720},
url = {https://www.sciencedirect.com/science/article/pii/S001046552500222X},
author = {Eva Kogler and Dominik Spath and Roman Lucrezi and Hitoshi Mori and Zien Zhu and Zhenglu Li and Elena R. Margine and Christoph Heil},
}

@article{Drozdov2015,
  title={Conventional superconductivity at 203 kelvin at high pressures in the sulfur hydride system},
  author={Drozdov, A. P. and Eremets, M. I. and Troyan, I. A. and Ksenofontov, V. and Shylin, S. I.},
  journal={Nature},
  volume={525},
  number={7567},
  pages={73--76},
  year={2015},
  publisher={Nature Publishing Group UK London},
  url = {https://doi.org/10.1038/nature14964},
  doi = {10.1038/nature14964}
}

@incollection{Allen1983,
title = {{Theory of Superconducting $T_{\rm c}$}},
editor = {Henry Ehrenreich and Frederick Seitz and David Turnbull},
series = {Solid State Physics},
publisher = {Academic Press},
volume = {37},
pages = {1-92},
year = {1983},
url = {https://www.sciencedirect.com/science/article/pii/S0081194708606657},
author = {Philip B. Allen and Božidar Mitrović}
}

@article{Lucrezi2024,
  title={{Full-bandwidth anisotropic Migdal-Eliashberg theory and its application to superhydrides}},
  author={Lucrezi, Roman and Ferreira, Pedro P and Hajinazar, Samad and Mori, Hitoshi and Paudyal, Hari and Margine, Elena R and Heil, Christoph},
  journal={Commun. Phys.},
  volume={7},
  number={1},
  pages={33},
  year={2024},
  publisher={Nature Publishing Group UK London},
  URL = {https://doi.org/10.1038/s42005-024-01528-6}
}

@article{Hamann2013,
  title = {{Optimized norm-conserving Vanderbilt pseudopotentials}},
  author = {Hamann, D. R.},
  journal = {Phys. Rev. B},
  volume = {88},
  issue = {8},
  pages = {085117},
  numpages = {10},
  year = {2013},
  month = {Aug},
  publisher = {American Physical Society},
  doi = {10.1103/PhysRevB.88.085117},
  url = {https://link.aps.org/doi/10.1103/PhysRevB.88.085117}
}

@article{Giannozzi2017,
  author = {Giannozzi, Paolo and Andreussi, Oliviero and Brumme, Thomas and Bunau, Olivier and Buongiorno Nardelli, Marco and Calandra, Matteo and Car, Roberto and Cavazzoni, Carlo and Ceresoli, Davide and Cococcioni, Matteo and Colonna, Nicola and Carnimeo, Ivan and Dal Corso, Andrea and de Gironcoli, Stefano and Delugas, Pietro and DiStasio Jr, Robert A. and Ferretti, Andrea and Floris, Andrea and Fratesi, Guido and Fugallo, Giorgio and Gebauer, Ralph and Gerstmann, Uwe and Giustino, Feliciano and Gorni, Tommaso and Jia, Jia and Kawamura, Masayuki and Ko, Hyo Yoon and Kokalj, Anton and Küçükbenli, Esen and Lazzeri, Michele and Marsili, Matteo and Marzari, Nicola and Mauri, Francesco and Nguyen, Ngoc Linh and Nguyen, Huy-Viet and Otero-de-la-Roza, Alberto and Paulatto, Lorenzo and Poncé, Samuel and Rocca, Dario and Sabatini, Riccardo and Santra, Biswajit and Schlipf, Martin and Seitsonen, Ari P. and Smogunov, Alexander and Timrov, Iurii and Thonhauser, Timo and Umari, Paolo and Vast, Nathalie and Wu, Xifan and Baroni, Stefano},
journal = {J. Phys.: Condens. Matter},
doi = {10.1088/1361-648x/aa8f79},
number = {46},
pages = {465901},
title = {{Advanced capabilities for materials modelling with Quantum ESPRESSO}},
volume = {29},
year = {2017}
}

@article{Perdew1996,
  title = {Generalized Gradient Approximation Made Simple},
  author = {Perdew, John P. and Burke, Kieron and Ernzerhof, Matthias},
  journal = {Phys. Rev. Lett.},
  volume = {77},
  issue = {18},
  pages = {3865--3868},
  numpages = {0},
  year = {1996},
  month = {Oct},
  publisher = {American Physical Society},
  doi = {10.1103/PhysRevLett.77.3865},
  url = {https://link.aps.org/doi/10.1103/PhysRevLett.77.3865}
}

@article{Marzari2012,
  title = {{Maximally localized Wannier functions: Theory and applications}},
  author = {Marzari, Nicola and Mostofi, Arash A. and Yates, Jonathan R. and Souza, Ivo and Vanderbilt, David},
  journal = {Rev. Mod. Phys.},
  volume = {84},
  issue = {4},
  pages = {1419--1475},
  numpages = {0},
  year = {2012},
  month = {Oct},
  publisher = {American Physical Society},
  doi = {10.1103/RevModPhys.84.1419},
  url = {https://link.aps.org/doi/10.1103/RevModPhys.84.1419}
}

@article{Pizzi2020,
	doi = {10.1088/1361-648x/ab51ff},
	url = {https://doi.org/10.1088%2F1361-648x%2Fab51ff},
	year = 2020,
	month = {jan},
	publisher = {{IOP} Publishing},
	volume = {32},
	number = {16},
	pages = {165902},
        author = {Pizzi, Giovanni and Vitale, Valerio and Arita, Ryotaro and Bl{\"u}gel, Stefan and Freimuth, Frank and G{\'e}ranton, Guillaume and Gibertini, Marco and Gresch, Dominik and Johnson, Charles and Koretsune, Takashi and Iba{\~n}ez-Azpiroz, Julen and Lee, Hyungjun and Lihm, Jae-Mo and Marchand, Daniel and Marrazzo, Antimo and Mokrousov, Yuriy and Mustafa, Jamal I. and Nohara, Yoshiro and Nomura, Yusuke and Paulatto, Lorenzo and Ponc{\'e}, Samuel and Ponweiser, Thomas and Qiao, Junfeng and Th{\"o}le, Florian and Tsirkin, Stepan S. and Wierzbowska, Ma{\l}gorzata and Marzari, Nicola and Vanderbilt, David and Souza, Ivo and Mostofi, Arash A. and Yates, Jonathan R.},
	title = {Wannier90 as a community code: new features and applications},
	journal = {J. Phys. Condens. Matter}
}

@article{Marazzo2024,
  title = {Wannier-function software ecosystem for materials simulations},
  author = {Marrazzo, Antimo and Beck, Sophie and Margine, Elena R. and Marzari, Nicola and Mostofi, Arash A. and Qiao, Junfeng and Souza, Ivo and Tsirkin, Stepan S. and Yates, Jonathan R. and Pizzi, Giovanni},
  journal = {Rev. Mod. Phys.},
  volume = {96},
  issue = {4},
  pages = {045008},
  numpages = {54},
  year = {2024},
  month = {Dec},
  publisher = {American Physical Society},
  doi = {10.1103/RevModPhys.96.045008},
  url = {https://link.aps.org/doi/10.1103/RevModPhys.96.045008}
}

@article{Baroni2001,
  title = {Phonons and related crystal properties from density-functional perturbation theory},
  author = {Baroni, Stefano and de Gironcoli, Stefano and Dal Corso, Andrea and Giannozzi, Paolo},
  journal = {Rev. Mod. Phys.},
  volume = {73},
  issue = {2},
  pages = {515--562},
  numpages = {0},
  year = {2001},
  month = {Jul},
  publisher = {American Physical Society},
  doi = {10.1103/RevModPhys.73.515},
  url = {https://link.aps.org/doi/10.1103/RevModPhys.73.515}
}

@article{Giustino2007,
  title = {{Electron-phonon interaction using Wannier functions}},
  author = {Giustino, Feliciano and Cohen, Marvin L. and Louie, Steven G.},
  journal = {Phys. Rev. B},
  volume = {76},
  issue = {16},
  pages = {165108},
  numpages = {19},
  year = {2007},
  month = {Oct},
  publisher = {American Physical Society},
  doi = {10.1103/PhysRevB.76.165108},
  url = {https://link.aps.org/doi/10.1103/PhysRevB.76.165108}
}

@article{Ponce2016,
title = {{{EPW}: Electron–phonon coupling, transport and superconducting properties using maximally localized Wannier functions}},
journal = {Comput. Phys. Commun.},
volume = {209},
pages = {116-133},
year = {2016},
issn = {0010-4655},
doi = {https://doi.org/10.1016/j.cpc.2016.07.028},
url = {https://www.sciencedirect.com/science/article/pii/S0010465516302260},
author = {S. Poncé and E.R. Margine and C. Verdi and F. Giustino}
}

@article{Margine2013,
  title = {{Anisotropic Migdal-Eliashberg theory using Wannier functions}},
  author = {Margine, E. R. and Giustino, F.},
  journal = {Phys. Rev. B},
  volume = {87},
  issue = {2},
  pages = {024505},
  numpages = {12},
  year = {2013},
  month = {Jan},
  publisher = {American Physical Society},
  doi = {10.1103/PhysRevB.87.024505},
  url = {https://link.aps.org/doi/10.1103/PhysRevB.87.024505}
}

@article{Sano2016,
  title = {{Effect of Van Hove singularities on high-${T}_{\mathrm{c}}$ superconductivity in ${\mathrm{H}}_{3}\mathrm{S}$}},
  author = {Sano, Wataru and Koretsune, Takashi and Tadano, Terumasa and Akashi, Ryosuke and Arita, Ryotaro},
  journal = {Phys. Rev. B},
  volume = {93},
  issue = {9},
  pages = {094525},
  numpages = {16},
  year = {2016},
  month = {Mar},
  publisher = {American Physical Society},
  doi = {10.1103/PhysRevB.93.094525},
  url = {https://link.aps.org/doi/10.1103/PhysRevB.93.094525}
}

@article{Gorkov2018,
  title = {Colloquium: High pressure and road to room temperature superconductivity},
  author = {Gor'kov, Lev P. and Kresin, Vladimir Z.},
  journal = {Rev. Mod. Phys.},
  volume = {90},
  issue = {1},
  pages = {011001},
  numpages = {16},
  year = {2018},
  month = {Jan},
  publisher = {American Physical Society},
  doi = {10.1103/RevModPhys.90.011001},
  url = {https://link.aps.org/doi/10.1103/RevModPhys.90.011001}
}

@article{Boeri2022,
	title = {The 2021 room-temperature superconductivity roadmap},
	volume = {34},
	issn = {0953-8984, 1361-648X},
	url = {https://iopscience.iop.org/article/10.1088/1361-648X/ac2864},
	doi = {10.1088/1361-648X/ac2864},
	journal = {J. Phys.: Condens. Matter},
	author = {Boeri, Lilia and Hennig, Richard and Hirschfeld, Peter and Profeta, Gianni and Sanna, Antonio and Zurek, Eva and Pickett, Warren E and Amsler, Maximilian and Dias, Ranga and Eremets, Mikhail I and Heil, Christoph and Hemley, Russell J and Liu, Hanyu and Ma, Yanming and Pierleoni, Carlo and Kolmogorov, Aleksey N and Rybin, Nikita and Novoselov, Dmitry and Anisimov, Vladimir and Oganov, Artem R and Pickard, Chris J and Bi, Tiange and Arita, Ryotaro and Errea, Ion and Pellegrini, Camilla and Requist, Ryan and Gross, E K U and Margine, Elena Roxana and Xie, Stephen R and Quan, Yundi and Hire, Ajinkya and Fanfarillo, Laura and Stewart, G R and Hamlin, J J and Stanev, Valentin and Gonnelli, Renato S and Piatti, Erik and Romanin, Davide and Daghero, Dario and Valenti, Roser},
	month = {may},
	year = {2022},
	pages = {183002},
}

@article{Monacelli2021,
doi = {10.1088/1361-648X/ac066b},
url = {https://dx.doi.org/10.1088/1361-648X/ac066b},
year = {2021},
month = {jul},
publisher = {IOP Publishing},
volume = {33},
number = {36},
pages = {363001},
author = {Lorenzo Monacelli and Raffaello Bianco and Marco Cherubini and Matteo Calandra and Ion Errea and Francesco Mauri},
title = {The stochastic self-consistent harmonic approximation: calculating vibrational properties of materials with full quantum and anharmonic effects},
journal = {J. Phys.: Condens. Matter}
}

@article{Gupta1978,
  title = {{Electronic structure and proton spin-lattice relaxation in PdH}},
  author = {Gupta, Mich\`ele and Freeman, A. J.},
  journal = {Phys. Rev. B},
  volume = {17},
  issue = {8},
  pages = {3029--3039},
  numpages = {0},
  year = {1978},
  month = {Apr},
  publisher = {American Physical Society},
  doi = {10.1103/PhysRevB.17.3029},
  url = {https://link.aps.org/doi/10.1103/PhysRevB.17.3029}
}

@article{Houari2014,
    author = {Houari, Abdesalem and Matar, Samir F. and Eyert, Volker},
    title = {Electronic structure and crystal phase stability of palladium hydrides},
    journal = {J. Appl. Phys.},
    volume = {116},
    number = {17},
    pages = {173706},
    year = {2014},
    month = {11},
    doi = {10.1063/1.4901004},
    url = {https://doi.org/10.1063/1.4901004}
}

@article{Worsham1957,
title = {Neutron-diffraction observations on the palladium-hydrogen and palladium-deuterium systems},
journal = {J. Phys. Chem. Solids},
volume = {3},
number = {3},
pages = {303-310},
year = {1957},
url = {https://www.sciencedirect.com/science/article/pii/0022369757900331},
author = {J.E. Worsham and M.K. Wilkinson and C.G. Shull},
}

@article{McLennan2008,
  title = {Deuterium occupation of tetrahedral sites in palladium},
  author = {McLennan, K. G. and Gray, E. MacA. and Dobson, J. F.},
  journal = {Phys. Rev. B},
  volume = {78},
  issue = {1},
  pages = {014104},
  numpages = {9},
  year = {2008},
  month = {Jul},
  publisher = {American Physical Society},
  doi = {10.1103/PhysRevB.78.014104},
  url = {https://link.aps.org/doi/10.1103/PhysRevB.78.014104}
}

@article{Talantsev2022,
    author = {Talantsev, Evgeny F.},
    title = {{Universal Fermi velocity in highly compressed hydride superconductors}},
    journal = {Matter Radiat. Extrem.},
    volume = {7},
    number = {5},
    pages = {058403},
    year = {2022},
    month = {08},
    issn = {2468-2047},
    doi = {10.1063/5.0091446},
    url = {https://doi.org/10.1063/5.0091446}
}

@article{Drozdov2019,
author={Drozdov, A. P. and Kong, P. P. and Minkov, V. S..and Besedin, S. P. and Kuzovnikov, M. A. and Mozaffari, S. and Balicas, L. and Balakirev, F. F. and Graf, D. E.
and Prakapenka, V. B. and Greenberg, E. and Knyazev, D. A. and Tkacz, M. and Eremets, M. I.},
title={{Superconductivity at 250 K in lanthanum hydride under high pressures}},
journal={Nature},
year={2019},
month={May},
day={01},
volume={569},
number={7757},
pages={528-531},
issn={1476-4687},
doi={10.1038/s41586-019-1201-8},
url={https://doi.org/10.1038/s41586-019-1201-8}
}

@article{Liu2017,
  title={{Potential high-T$_{\rm c}$ superconducting lanthanum and yttrium hydrides at high pressure}},
  author={Liu, Hanyu and Naumov, Ivan I and Hoffmann, Roald and Ashcroft, NW and Hemley, Russell J},
  journal={Proc. Natl. Acad.Sci.},
  volume={114},
  number={27},
  pages={6990--6995},
  year={2017},
  publisher={National Acad Sciences},
  url = {www.pnas.org/cgi/doi/10.1073/pnas.1704505114}
}

@article{Somayazulu2019,
  title = {{Evidence for Superconductivity above 260 K in Lanthanum Superhydride at Megabar Pressures}},
  author = {Somayazulu, Maddury and Ahart, Muhtar and Mishra, Ajay K. and Geballe, Zachary M. and Baldini, Maria and Meng, Yue and Struzhkin, Viktor V. and Hemley, Russell J.},
  journal = {Phys. Rev. Lett.},
  volume = {122},
  issue = {2},
  pages = {027001},
  numpages = {6},
  year = {2019},
  month = {Jan},
  publisher = {American Physical Society},
  doi = {10.1103/PhysRevLett.122.027001},
  url = {https://link.aps.org/doi/10.1103/PhysRevLett.122.027001}
}

@article{Peng2017,
  title = {{Hydrogen Clathrate Structures in Rare Earth Hydrides at High Pressures: Possible Route to Room-Temperature Superconductivity}},
  author = {Peng, Feng and Sun, Ying and Pickard, Chris J. and Needs, Richard J. and Wu, Qiang and Ma, Yanming},
  journal = {Phys. Rev. Lett.},
  volume = {119},
  issue = {10},
  pages = {107001},
  numpages = {6},
  year = {2017},
  month = {Sep},
  publisher = {American Physical Society},
  doi = {10.1103/PhysRevLett.119.107001},
  url = {https://link.aps.org/doi/10.1103/PhysRevLett.119.107001}
}

@article{Ashcroft2004,
  title = {{Hydrogen Dominant Metallic Alloys: High Temperature Superconductors?}},
  author = {Ashcroft, N. W.},
  journal = {Phys. Rev. Lett.},
  volume = {92},
  issue = {18},
  pages = {187002},
  numpages = {4},
  year = {2004},
  month = {May},
  publisher = {American Physical Society},
  doi = {10.1103/PhysRevLett.92.187002},
  url = {https://link.aps.org/doi/10.1103/PhysRevLett.92.187002}
}

@article{Zurek2019,
    author = {Zurek, Eva and Bi, Tiange},
    title = {{High-temperature superconductivity in alkaline and rare earth polyhydrides at high pressure: A theoretical perspective}},
    journal = {J. Chem. Phys.},
    volume = {150},
    number = {5},
    pages = {050901},
    year = {2019},
    month = {02},
    doi = {10.1063/1.5079225},
    url = {https://doi.org/10.1063/1.5079225}
}

@article{Ding2025,
  title = {{Anharmonicity and Coulomb pseudopotential effects on superconductivity in ${\mathrm{YH}}_{6}$ and ${\mathrm{YH}}_{9}$}},
  author = {Ding, Yucheng and Chen, Haoran and Shi, Junren},
  journal = {Phys. Rev. B},
  volume = {112},
  issue = {18},
  pages = {184517},
  numpages = {7},
  year = {2025},
  month = {Nov},
  publisher = {American Physical Society},
  doi = {10.1103/8y74-91v2},
  url = {https://link.aps.org/doi/10.1103/8y74-91v2}
}

@article{Wang2022,
doi = {10.1088/1674-1056/ac872e},
url = {https://doi.org/10.1088/1674-1056/ac872e},
year = {2022},
month = {sep},
publisher = {Chinese Physical Society and IOP Publishing Ltd},
volume = {31},
number = {10},
pages = {106201},
author = {Wang, Yingying and Wang, Kui and Sun, Yao and Ma, Liang and Wang, Yanchao and Zou, Bo and Liu, Guangtao and Zhou, Mi and Wang, Hongbo},
title = {Synthesis and superconductivity in yttrium superhydrides under high pressure},
journal = {Chin. Phys. B}
}

@article{Li2015,
  title={Pressure-stabilized superconductive yttrium hydrides},
  author={Li, Yinwei and Hao, Jian and Liu, Hanyu and Tse, John S and Wang, Yanchao and Ma, Yanming},
  journal={Sci. Rep.},
  volume={5},
  number={1},
  pages={9948},
  year={2015},
  publisher={Nature Publishing Group UK London},
  url={https://doi.org/10.1038/srep09948}
}

@article{Dong2025,
title = {{Machine-learning potentials for quantum and anharmonic effects in superconducting Fm$\bar{3}$m ${\mathrm{LaBeH}}_{8}$}},
journal = {Mater. Today Phys.},
volume = {59},
pages = {101939},
year = {2025},
issn = {2542-5293},
doi = {https://doi.org/10.1016/j.mtphys.2025.101939},
url = {https://www.sciencedirect.com/science/article/pii/S2542529325002950},
author = {Guiyan Dong and Tian Cui and Zihao Huo and Zhengtao Liu and Wenxuan Chen and Pugeng Hou and Yue-Wen Fang and Defang Duan}
}

@article{Zhang2022,
  title = {{Design Principles for High-Temperature Superconductors with a Hydrogen-Based Alloy Backbone at Moderate Pressure}},
  author = {Zhang, Zihan and Cui, Tian and Hutcheon, Michael J. and Shipley, Alice M. and Song, Hao and Du, Mingyang and Kresin, Vladimir Z. and Duan, Defang and Pickard, Chris J. and Yao, Yansun},
  journal = {Phys. Rev. Lett.},
  volume = {128},
  issue = {4},
  pages = {047001},
  numpages = {7},
  year = {2022},
  month = {Jan},
  publisher = {American Physical Society},
  doi = {10.1103/PhysRevLett.128.047001},
  url = {https://link.aps.org/doi/10.1103/PhysRevLett.128.047001}
}

@article{Rowe1986,
  title = {{Isotope Effects in the PdH System: Lattice Dynamics of Pd${\mathrm{T}}_{0.7}$}},
  author = {Rowe, J. M. and Rush, J. J. and Schirber, J. E. and Mintz, J. M.},
  journal = {Phys. Rev. Lett.},
  volume = {57},
  issue = {23},
  pages = {2955--2958},
  numpages = {0},
  year = {1986},
  month = {Dec},
  publisher = {American Physical Society},
  doi = {10.1103/PhysRevLett.57.2955},
  url = {https://link.aps.org/doi/10.1103/PhysRevLett.57.2955}
}

@article{Ganguly1976,
  title = {{Anharmonicity and superconductivity in metal hydrides. I. Formulation}},
  author = {Ganguly, B. N.},
  journal = {Phys. Rev. B},
  volume = {14},
  issue = {9},
  pages = {3848--3852},
  numpages = {0},
  year = {1976},
  month = {Nov},
  publisher = {American Physical Society},
  doi = {10.1103/PhysRevB.14.3848},
  url = {https://link.aps.org/doi/10.1103/PhysRevB.14.3848}
}

@article{Schirber1984,
title = {Superconductivity of palladium tritide},
journal = {Solid State Commun.},
volume = {52},
number = {10},
pages = {837-838},
year = {1984},
issn = {0038-1098},
doi = {https://doi.org/10.1016/0038-1098(84)90251-5},
url = {https://www.sciencedirect.com/science/article/pii/0038109884902515},
author = {J.E. Schirber and J.M. Mintz and W. Wall},
}

@article{Papaconstantopoulos1978,
  title = {{Band structure and superconductivity of $\mathrm{Pd}{\mathrm{D}}_{x}$ and $\mathrm{Pd}{\mathrm{H}}_{x}$}},
  author = {Papaconstantopoulos, D. A. and Klein, B. M. and Economou, E. N. and Boyer, L. L.},
  journal = {Phys. Rev. B},
  volume = {17},
  issue = {1},
  pages = {141--150},
  numpages = {0},
  year = {1978},
  month = {Jan},
  publisher = {American Physical Society},
  doi = {10.1103/PhysRevB.17.141},
  url = {https://link.aps.org/doi/10.1103/PhysRevB.17.141}
}

@article{Jena1984,
  title = {{Effect of zero-point motion on the superconducting transition temperature of PdH(D)}},
  author = {Jena, P. and Jones, J. and Nieminen, R. M.},
  journal = {Phys. Rev. B},
  volume = {29},
  issue = {7},
  pages = {4140--4143},
  numpages = {0},
  year = {1984},
  month = {Apr},
  publisher = {American Physical Society},
  doi = {10.1103/PhysRevB.29.4140},
  url = {https://link.aps.org/doi/10.1103/PhysRevB.29.4140}
}

@article{Rowe1974,
  title = {{Lattice Dynamics of a Single Crystal of Pd${\mathrm{D}}_{0.63}$}},
  author = {Rowe, J. M. and Rush, J. J. and Smith, H. G. and Mostoller, Mark and Flotow, H. E.},
  journal = {Phys. Rev. Lett.},
  volume = {33},
  issue = {21},
  pages = {1297--1300},
  numpages = {0},
  year = {1974},
  month = {Nov},
  publisher = {American Physical Society},
  doi = {10.1103/PhysRevLett.33.1297},
  url = {https://link.aps.org/doi/10.1103/PhysRevLett.33.1297}
}

@article{Miller1975,
  title = {{Electronic Model for the Reverse Isotope Effect in Superconducting Pd-H(D)}},
  author = {Miller, R. J. and Satterthwaite, C. B.},
  journal = {Phys. Rev. Lett.},
  volume = {34},
  issue = {3},
  pages = {144--148},
  numpages = {0},
  year = {1975},
  month = {Jan},
  publisher = {American Physical Society},
  doi = {10.1103/PhysRevLett.34.144},
  url = {https://link.aps.org/doi/10.1103/PhysRevLett.34.144}
}

@article{Liu2025,
  title = {{Impact of deviatoric stress on the stability and superconductivity of ${\mathrm{H}}_{3}\mathrm{S}$}},
  author = {Liu, Han and Liu, Chang and Ma, Yanming and Chen, Changfeng},
  journal = {Phys. Rev. B},
  volume = {112},
  issue = {13},
  pages = {134518},
  numpages = {12},
  year = {2025},
  month = {Oct},
  publisher = {American Physical Society},
  doi = {10.1103/sn9r-3cx2},
  url = {https://link.aps.org/doi/10.1103/sn9r-3cx2}
}

@article{Mishra2025,
  author={Shashi B. Mishra and Hitoshi Mori and Elena R. Margine},
  title = {{Electron-phonon vertex correction effect in superconducting H$_3$S}}, 
  journal = {npj Comput. Mater.},
  volume = {11},
  issue = {1},
  pages = {342},
  year = {2025},
  doi={10.1038/s41524-025-01818-9},
  url={https://doi.org/10.1038/s41524-025-01818-9}
}

@article{Belli2025,
  title={{Efficient modelling of anharmonicity and quantum effects in PdCuH$_2$ with machine learning potentials}},
  author={Belli, Francesco and Zurek, Eva},
  journal={npj Comput. Mater.},
  volume={11},
  number={1},
  pages={87},
  year={2025},
  publisher={Nature Publishing Group UK London},
  url={https://www.nature.com/articles/s41524-025-01553-1}
}

@article{Skoskiewicz1972,
  title = {Superconductivity in the palladium-hydrogen and palladium-nickel-hydrogen systems},
  author = {Sko\'{s}kiewicz, T.},
  journal = {Phys. Status Solidi A},
  volume = {11},
  number = {2},
  pages = {K123--K126},
  year = {1972},
  doi = {https://doi.org/10.1002/pssa.2210110253}
}

@article{Yussouff1995,
title = {{Reverse isotope effect on the superconductivity of PdH, PdD, and PdT}},
journal = {Solid State Commun.},
volume = {94},
number = {7},
pages = {549-553},
year = {1995},
issn = {0038-1098},
doi = {https://doi.org/10.1016/0038-1098(94)00909-0},
url = {https://www.sciencedirect.com/science/article/pii/0038109894009090},
author = {M. Yussouff and B.K. Rao and P. Jena}
}

@article{Stritzker1972,
  title = {Superconductivity in the palladium-hydrogen and the palladium-deuterium systems},
  author = {Stritzker, B. and Buckel, W.},
  journal = {Z. Phys.},
  volume = {257},
  number = {1},
  pages = {1--8},
  year = {1972},
  doi = {https://doi.org/10.1007/BF01398191}
}

@article{Schirber1974,
  title = {{Concentration dependence of the superconducting transition temperature in $\mathrm{PdH}_x$ and $\mathrm{PdD}_x$}},
  author = {Schirber, J. E. and Northrup, C. J. M.},
  journal = {Phys. Rev. B},
  volume = {10},
  number = {9},
  pages = {3818--3820},
  year = {1974},
  doi = {https://doi.org/10.1103/PhysRevB.10.3818}
}

@article{Griessen1982,
  title = {{Effect of anharmonicity and Debye-Waller factor on the superconductivity of PdH$_x$ and PdD$_x$}},
  author = {Griessen, R. and Groot, D.G. de},
  journal = {Helv. Phys. Acta},
  volume = {55},
  number = {6},
  pages = {699},
  year = {1982},
  doi = {https://doi.org/10.5169/seals-115306}
}

@article{Wang2024,
  title = {Unlocking the origin of stability and superconductivity in {LaBeH$_8$} at submegabar pressure},
  author = {Wang, Zefang and Zhao, Hongjian and Zhong, Xin and Liu, Hanyu and Ma, Yanming},
  journal = {Phys. Rev. B},
  volume = {109},
  pages = {214506},
  year = {2024},
  doi = {10.1103/PhysRevB.109.214506}
}

\end{document}


\title{Supplemental Material: \\ Anharmonicity and Nonadiabaticity in Hydride Superconductors}

\author{Shashi B. Mishra}
\email{mshashi125@gmail.com}
\affiliation{Department of Physics, Applied Physics and Astronomy, Binghamton University-SUNY, Binghamton, New York 13902, USA}
\author{Francesco Belli}
\affiliation{Department of Chemistry, State University of New York at Buffalo, Buffalo, NY, USA}
\author{Eva Zurek}
\affiliation{Department of Chemistry, State University of New York at Buffalo, Buffalo, NY, USA}
\author{Elena R. Margine}
\email{rmargine@binghamton.edu}
\affiliation{Department of Physics, Applied Physics and Astronomy, Binghamton University-SUNY, Binghamton, New York 13902, USA}
\date{\today}

\maketitle
\tableofcontents

\section{Computational workflow}
\label{sec:workflow}

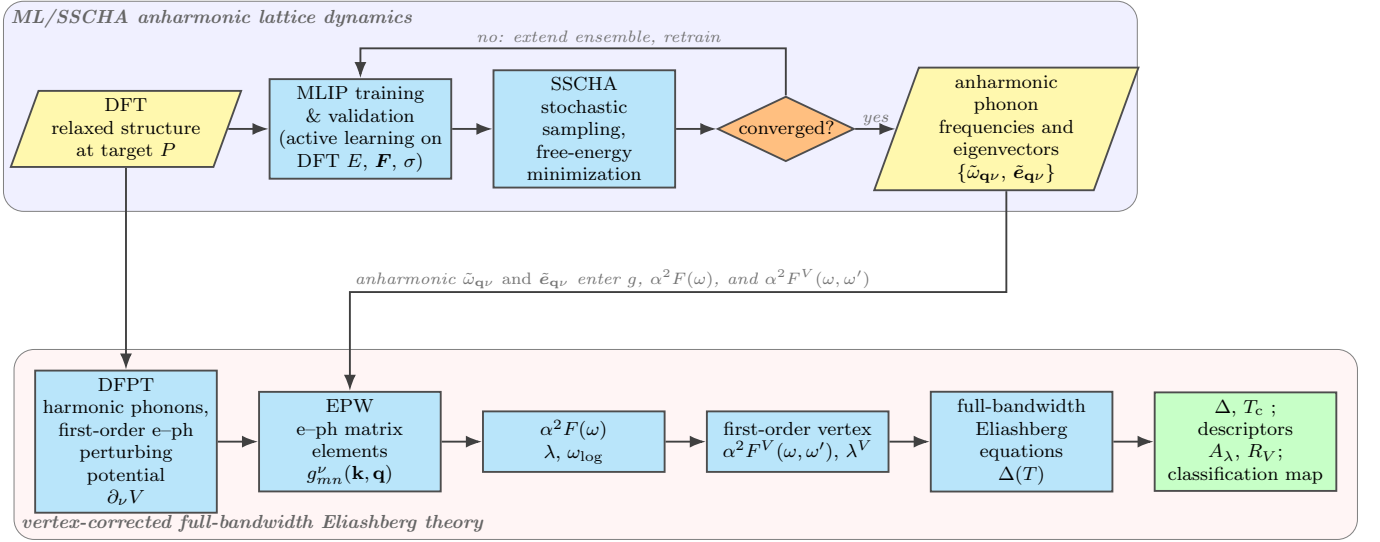
\begin{figure*}[!t]
\centering
\resizebox{\textwidth}{!}{%
\begin{tikzpicture}[
  font=\footnotesize,
  node distance=4.5mm and 6mm,
  io/.style={trapezium, trapezium left angle=70, trapezium right angle=110,
             draw=black!70, thick, fill=yellow!40, minimum height=9mm,
             align=center, inner xsep=1.5mm, text width=23mm},
  proc/.style={rectangle, draw=black!70, thick, fill=cyan!25,
               minimum height=9mm, align=center, inner xsep=1mm, text width=25mm},
  dec/.style={diamond, aspect=2.1, draw=black!70, thick, fill=orange!50,
              align=center, inner sep=0.5pt, text width=14mm},
  res/.style={rectangle, draw=black!70, thick, fill=green!22,
              minimum height=9mm, align=center, inner xsep=1mm, text width=25mm},
  arr/.style={-{Latex[length=2.4mm]}, thick, draw=black!75},
  lbl/.style={font=\scriptsize\itshape, text=black!55, align=center, inner sep=1pt},
]
\node[io]                        (dft)   {DFT\\ relaxed structure\\ at target $P$};
\node[proc, right=of dft]        (mlip)  {MLIP training\\ \& validation\\ (active learning on\\ DFT $E$, $\bm{F}$, $\sigma$)};
\node[proc, right=of mlip]       (sscha) {SSCHA\\ stochastic sampling,\\ free-energy\\ minimization};
\node[dec,  right=of sscha]      (conv)  {converged?};
\node[io,   right=of conv]       (adyn)  {anharmonic phonon\\ frequencies and\\ eigenvectors\\ $\{\tilde{\o}_{\bq\nu},\,\tilde{\bm{e}}_{\bq\nu}\}$};
\draw[arr] (dft)   -- (mlip);
\draw[arr] (mlip)  -- (sscha);
\draw[arr] (sscha) -- (conv);
\draw[arr] (conv)  -- node[lbl, above] {yes} (adyn);
\draw[arr] (conv.north) -- ++(0,7mm) -| node[lbl, above, pos=0.22] {no: extend ensemble, retrain} (mlip.north);
\coordinate (fitpad) at ($(conv.north)+(0,11mm)$);
\node[proc, below=30mm of dft]   (dfpt)  {DFPT\\ harmonic phonons,\\ first-order e--ph\\ perturbing potential\\ $\partial_{\nu}V$};
\node[proc, right=of dfpt]       (wan)   {EPW\\ e--ph matrix elements\\ $g_{mn}^{\nu}(\bk,\bq)$};
\node[proc, right=of wan]        (a2f)   {\aaff{}\\ $\l$, \wlg{}};
\node[proc, right=of a2f]        (vtx)   {first-order vertex\\ \aaffv{}, $\lV$};
\node[proc, right=of vtx]        (eli)   {full-bandwidth\\ Eliashberg\\ equations\\ $\Delta(T)$};
\node[res,  right=of eli]        (out)   {$\Delta$, $\tc$ ;\\ descriptors $\AL$, $\RV$;\\ classification map};
\draw[arr] (dft)  -- (dfpt);
\draw[arr] (dfpt) -- (wan);
\draw[arr] (wan)  -- (a2f);
\draw[arr] (a2f)  -- (vtx);
\draw[arr] (vtx)  -- (eli);
\draw[arr] (eli)  -- (out);
\draw[arr] (adyn.south) -- ++(0,-15mm) -| node[lbl, above, pos=0.3]
  {anharmonic $\tilde{\o}_{\bq\nu}$ \emph{and} $\tilde{\bm{e}}_{\bq\nu}$ enter $g$, \aaff{}, and \aaffv{}} (wan.north);
\begin{scope}[on background layer]
\node[rectangle, rounded corners=3mm, draw=black!50, fill=blue!6, inner sep=3mm,
      fit=(dft)(mlip)(sscha)(conv)(adyn)(fitpad), label={[anchor=north west, font=\footnotesize\bfseries\itshape, text=black!60]north west:ML/SSCHA anharmonic lattice dynamics}] (box1) {};
\coordinate (fitpad2) at ($(wan.south)+(0,-4mm)$);
\node[rectangle, rounded corners=3mm, draw=black!50, fill=red!4, inner sep=3mm,
      fit=(dfpt)(wan)(a2f)(vtx)(eli)(out)(fitpad2), label={[anchor=south west, font=\footnotesize\bfseries\itshape, text=black!60]south west:vertex-corrected full-bandwidth Eliashberg theory}] (box2) {};
\end{scope}
\end{tikzpicture}}
\caption{First-principles workflow for anharmonic and vertex-corrected superconductivity calculations. Yellow parallelograms denote input and output data, blue rectangles computational steps, the orange diamond a convergence check, and the green box the final observables. The ML/SSCHA block provides anharmonic phonon frequencies and eigenvectors~\cite{Belli2025}, which enter the EPW evaluation of the electron-phonon matrix elements, \aaff{}, and \aaffv{}~\cite{Mishra2025}.}
\label{fig:workflow}
\end{figure*}

Figure~\ref{fig:workflow} summarizes the computational workflow. 
Harmonic phonons and first-order electron-phonon (e-ph) perturbing potentials are obtained from density-functional perturbation theory (DFPT)~\cite{Baroni2001}. The corresponding e-ph matrix elements are constructed and Wannier interpolated with EPW~\cite{Giustino2007}. Anharmonic lattice dynamics are treated with the machine-learning-accelerated stochastic self-consistent harmonic approximation (ML/SSCHA)~\cite{Belli2025,Monacelli2021}. The density-functional theory (DFT) energies, forces, and stresses are used to train an interatomic potential within an active-learning protocol. The converged auxiliary force constants yield the anharmonic phonon frequencies $\tilde{\o}_{\bq\nu}$ and eigenvectors $\tilde{\bm{e}}_{\bq\nu}$. These quantities are combined with the linear e-ph matrix elements to construct the Eliashberg spectral function \aaff{} and the two-frequency vertex spectral function \aaffv{}, whose weighted double integral defines the nonadiabatic coupling $\lV$~\cite{Mishra2025}. Next, the full-bandwidth (FBW) Eliashberg equations are solved with and without first-order vertex correction, yielding $\Delta(T)$ and $\tc$. Anharmonicity therefore enters through the renormalized phonon frequencies and eigenvectors, whereas nonadiabaticity enters through the first-order vertex kernel encoded in \aaffv{}. Finally, the descriptors $\AL$ and $\RV$ are evaluated from the calculated quantities as a post-processing step and used solely to classify the materials.

\section{Computational details}
\label{sec:comp}
Electronic-structure calculations were performed with Quantum~{\small ESPRESSO}~\cite{Giannozzi2017} using optimized norm-conserving Vanderbilt pseudopotentials~\cite{Hamann2013} generated with the Perdew-Burke-Ernzerhof (PBE) exchange-correlation functional~\cite{Perdew1996}. Plane-wave kinetic-energy cutoffs of 60~Ry (H$_3$S), 70~Ry (YH$_6$, YH$_9$, LaBeH$_8$), and 80~Ry (Pd hydrides) were used with Methfessel-Paxton smearing of 0.01~Ry for high-pressure hydrides (H$_3$S, YH$_6$, YH$_9$, and LaBeH$_8$), and 0.02~Ry for the ambient pressure Pd hydrides. Lattice parameters and internal coordinates were relaxed at the target static-lattice pressure until forces were below $10^{-4}$~Ry/\AA{} and total energies were converged to $10^{-6}$~Ry. The corresponding quantum-corrected pressure $\tilde{P}$, obtained by adding the SSCHA quantum-nuclear contribution to the stress of the same cell, is listed in Table~\ref{tab:structures}. Depending on the hydride, this correction increases the pressure by approximately 10--20~GPa. For YH$_9$, a subsequent relaxation was performed through the SSCHA at its corresponding quantum-corrected pressure. Dense $\Gamma$-centered electronic $k$-grids were used for the ground state and density of states (DOS): $24\times24\times24$ for H$_3$S and YH$_6$, $25\times25\times25$ for the Pd hydrides, $12\times12\times12$ for LaBeH$_8$, and $12\times12\times8$ for YH$_9$. PdD and PdT share the static lattice, and therefore the electronic structure, of PdH; the isotope mass enters only through the lattice dynamics.

\begin{table}[!hbt]
\caption{Structures used in the calculations. Starting from the experimentally assigned phases cited in the last column, all structures were relaxed at the target static-lattice pressures $P$, while $\tilde{P}$ reports the pressure with the inclusion of quantum nuclear effects treated with the SSCHA. The optimized conventional-cell lattice parameters at the target static-lattice pressure are listed for comparison with previous studies.} 
\label{tab:structures}
\begin{ruledtabular}
\begin{tabular}{l c c c c l l}
Material & $P$ (GPa) &  $\tilde{P}$ (GPa) & Space group & Lattice parameter (\AA) & Structural motif & Phase ref. \\
\colrule
H$_3$S       & 200 &  210 & $Im\bar{3}m$  & $a=2.986$ & covalent S--H network        & \cite{Drozdov2015} \\
YH$_6$       & 166 &  180 & $Im\bar{3}m$  & $a=3.576$ & Y-stabilized sodalite H network         & \cite{Troyan2021} \\
YH$_9$       & 231 &  250 & $P6_3/mmc$    & $a=3.342$, $c=5.190$ & Y-stabilized clathrate H network       & \cite{Kong2021} \\
LaBeH$_8$    &  81 &  90  & $Fm\bar{3}m$  & $a=5.446$ & fluorite-type, BeH$_8$ units  & \cite{Song2023} \\
PdH/D/T      &   0 &  -   & $Fm\bar{3}m$  & $a=4.093$ & rocksalt interstitial H      & \cite{Skoskiewicz1972} \\
\end{tabular}
\end{ruledtabular}
\end{table}

Harmonic dynamical matrices and the linear variation of the self-consistent potential were obtained from DFPT~\cite{Baroni2001,Giannozzi2017} on uniform $4\times4\times4$ $q$-meshes for all systems except YH$_9$, for which a $4\times4\times2$ mesh was used. Anharmonic dynamical matrices were computed using the ML/SSCHA workflow~\cite{Belli2025} (Figure~\ref{fig:workflow}, top), with the same procedure applied to all systems. For the machine learning workflow, the anharmonic-quantum dynamical matrices were obtained by training an 18-level Moment Tensor Potential on a $2\times2\times2$ supercell for H$_3$S, YH$_6$, PdH/D/T, LaBeH$_8$, and a $2\times2\times1$ supercell for YH$_9$ with 32, 56, 16, 80, and 160 atoms, respectively. These calculations were performed allowing the atomic positions to relax at fixed pressure for YH$_9$, and allowing the atomic positions to relax at fixed cell volume for H$_3$S, YH$_6$, LaBeH$_8$, and PdH/D/T. The SSCHA calculations were progressively repeated in larger supercells, reaching $4\times4\times4$ for all systems except YH$_9$, for which a $4\times4\times2$ supercell was used. The bubble approximation was used for the final dynamical matrices. 

Using EPW~\cite{Lee2023}, we constructed and Wannier interpolated the e-ph matrix elements~\cite{Giustino2007,Giustino2017,Marzari2012,Marazzo2024} and computed the isotropic Eliashberg spectral function \aaff{}, the e-ph coupling $\l=2\int\aaff{}\o^{-1}d\o$, and the logarithmic average frequency $\wlg$~\cite{Margine2013,Ponce2016}. The two-frequency vertex spectral function \aaffv{} and the corresponding nonadiabatic coupling $\lV$ were evaluated using the first-order vertex implementation of Ref.~\cite{Mishra2025}. Starting from coarse $8\times8\times8$ electronic $k$- and $4\times4\times4$ phonon $q$-grids for all systems except YH$_9$, for which $8\times8\times4$ and $4\times4\times2$ grids were used, the matrix elements were interpolated onto dense fine meshes: $48\times48\times48$~($k$)/$24\times24\times24$~($q$) for H$_3$S, YH$_6$, and the Pd hydrides; $40\times40\times40$/$20\times20\times20$ for LaBeH$_8$; and $36\times36\times24$/$18\times18\times12$ for YH$_9$. Electronic states within $\pm0.2$~eV of $\ef$ were included.
The isotropic Eliashberg equations were solved on the Matsubara frequency axis using full-bandwidth (FBW, energy-dependent DOS) formalisms~\cite{Lee2023,Lucrezi2024,Mishra2025}, using a Matsubara frequency cutoff of about ten times the maximum phonon frequency of each system (2.3--2.6~eV for the high-pressure hydrides and 0.7--1.2~eV for the Pd hydrides). For all high-pressure hydrides, we chose the conventional value $\mu^{*}=0.13$, commonly adopted for such systems~\cite{Liu2017,Peng2017,Liu2025, Cataldo2021, Zurek2019}, whereas the lower value $\mu^{*}=0.085$ was applied to the ambient-pressure Pd hydrides for consistency with the anharmonic SSCHA benchmark of Ref.~\cite{Errea2013}. The sensitivity of $\tc$ to $\mu^{*}$ is analyzed in Sec.~\ref{sec:supercond}.

\section{Electronic and vibrational properties}
\label{sec:elec}

\begin{figure}[!hbt]
    \centering
    \includegraphics[width=0.9\textwidth]{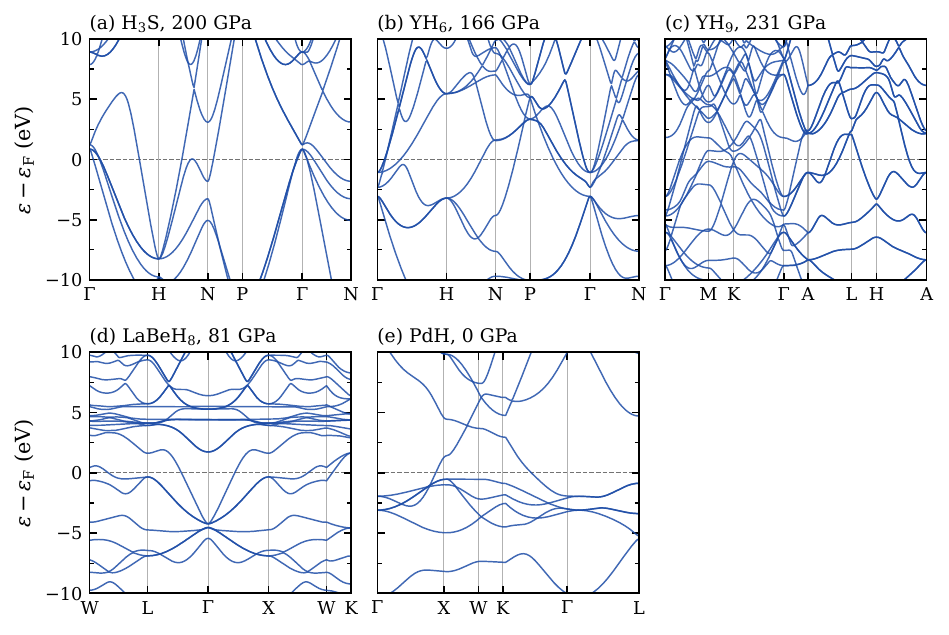}
    \caption{Electronic band structures for (a)~H$_3$S, (b)~YH$_6$, (c)~YH$_9$, (d)~LaBeH$_8$, and (e)~PdH along their respective high-symmetry paths, with corresponding static-lattice pressure $P$ indicated. Energies are referenced to the Fermi level $\ef=0$ (dashed line).}
    \label{fig:bands}
\end{figure}

\begin{figure}[!hbt]
    \centering
    \includegraphics[width=0.9\textwidth]{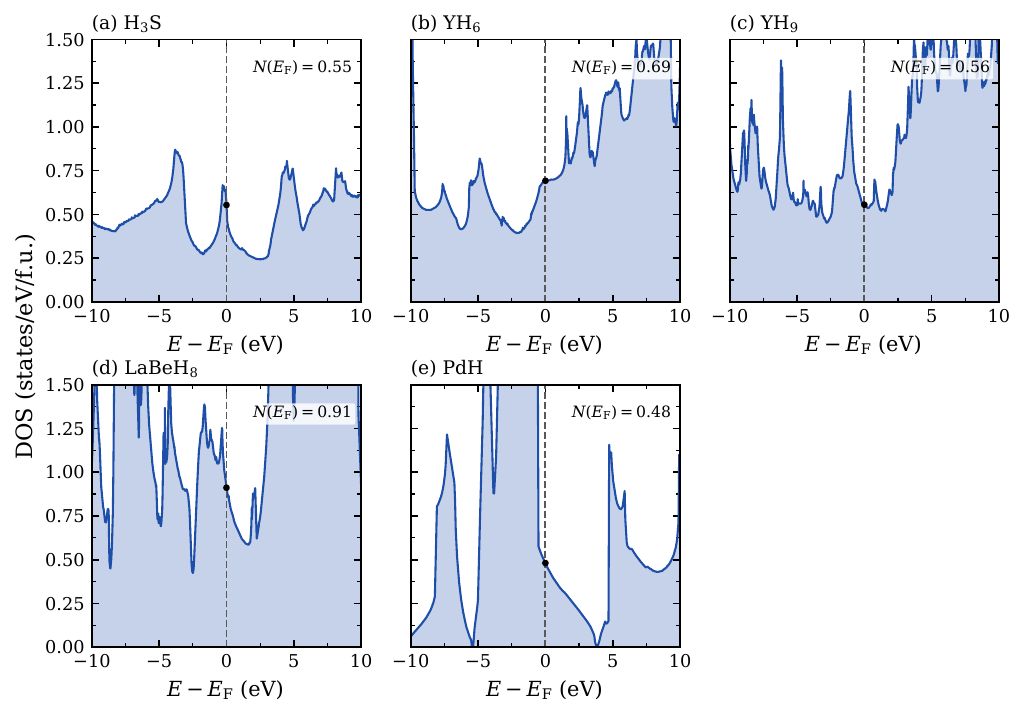}
    \caption{Total electronic DOS per formula unit for (a)~H$_3$S, (b)~YH$_6$, (c)~YH$_9$, (d)~LaBeH$_8$, and (e)~PdH. The DOS at the Fermi level, $N(E_F)$, is marked by a solid dot and labeled in each panel; $\ef$ is indicated by the dashed vertical line.} 
    \label{fig:dos}
\end{figure}

Figure~\ref{fig:bands} shows the electronic band structures for H$_3$S, YH$_6$, YH$_9$, LaBeH$_8$, and PdH, while Fig.~\ref{fig:dos} shows the corresponding total DOS. The onsite energies and hopping amplitudes listed in Table~\ref{tab:hoppings} are obtained from Wannier Hamiltonians~\cite{Marzari2012,Pizzi2020}. The onsite energies are measured relative to the Fermi level ($\ef$), and each hopping is reported together with the distance between the corresponding Wannier centers.
In $Im\bar{3}m$ H$_3$S, the H~$1s$ and S~$3p$ levels are close in energy and are strongly hybridized. The S--H and H--H hopping amplitudes reach 5.5 and 1.8~eV, respectively. These large hopping amplitudes produce broad covalent H-$s$/S-$p$ bands at $\ef$. The sharp DOS peak just below $\ef$ has been associated with an extended saddle feature in the band structure~\cite{Akashi2020}. Consequently, the DOS varies rapidly over the energy range sampled by the high-frequency H modes, making H$_3$S particularly sensitive to full-bandwidth and vertex corrections.
In YH$_6$, the hydrogen cage forms the primary conducting network, with H--H hopping amplitudes up to 3.5~eV exceeding the Y-$d$--H-$s$ channel ($\sim$1.9~eV). Hybridization nevertheless gives the states at $\ef$ mixed Y-$d$/H-$s$ character. Although the DOS is smoother than in H$_3$S, its variation across the pairing window makes full-bandwidth effects relevant.
YH$_9$ is more strongly hydrogen-dominated. Its shorter H--H contacts give hopping amplitudes up to 4.7~eV, and the states at $\ef$ are mainly H-$s$-like. Here, $\ef$ lies in a DOS minimum between nearby peaks. 
In LaBeH$_8$, the H-derived bonding states of the BeH$_8$ units lie well below $\ef$, while the La-$d$ level lies several eV above it. The H--H and La-$d$--H-$s$ hopping amplitudes are also smaller than in H$_3$S and the Y hydrides. The resulting DOS varies relatively gradually near $\ef$, which lies on the descending edge of a broad shoulder.
In PdH, the Pd-$d$/H-$s$ states lie below $\ef$, while a broad $s$-like band crosses the Fermi level. The corresponding Wannier function is centered at the tetrahedral interstitial sites and couples strongly to H-$s$, with a hopping amplitude of 2.8~eV at 1.77~\AA. Direct H--H hopping is negligible because the H atoms are much farther apart than in the clathrate hydrides. The resulting DOS varies smoothly and approximately linearly near $\ef$. 

\begin{table*}[!hbt]
\caption{Real-space Wannier analysis of the electronic structure. For each material, we report the Wannier basis, the mean H-$s$ and non-H (metal M) onsite energies relative to $\ef$, and the largest H--H, M--H, and M--M hopping amplitudes $|t|$ with their Wannier-center distances $d$.}
\label{tab:hoppings}
\begin{ruledtabular}
\begin{tabular}{l l c c c c c c c c}
Material & Basis & $\bar\varepsilon_{\rm H}-\ef$ & $\bar\varepsilon_{M}-\ef$ & \multicolumn{2}{c}{H--H} & \multicolumn{2}{c}{M--H} & \multicolumn{2}{c}{M--M} \\
    & & (eV) & (eV) & $|t|$ (eV) & $d$ (\AA) & $|t|$ (eV) & $d$ (\AA) & $|t|$ (eV) & $d$ (\AA) \\
\colrule
H$_3$S     & H $s$; S $s,p,d_{xy},d_{xz},d_{yz}$ 
           & $-1.4$ & $-0.8$ (S $p$) & $1.80$ & $1.49$ & $5.55$ & $1.49$ & $1.39$ & $2.99$ \\

YH$_6$     & Y $d$; H $s$
           & $-1.8$ & $+3.8$ (Y $d$) & $3.51$ & $1.24$ & $1.88$ & $2.02$ & $0.65$ & $3.10$ \\

YH$_9$     & Y $d$; H $s$
           & $-1.1$ & $+4.3$ (Y $d$) & $4.74$ & $1.03$ & $2.41$ & $1.92$ & $0.43$ & $3.22$ \\

LaBeH$_8$  & La $d$; H $s$
           & $-3.1$ & $+4.4$ (La $d$) & $2.08$ & $1.53$ & $2.23$ & $1.94$ & $0.25$ & $3.44$ \\

PdH        & Pd $d,s$; H $s$
           & $-2.1$ & $-2.1$ (Pd $d$) & $0.17$ & $4.09$ & $1.60$ & $2.05$ & $0.52$ & $2.73$ \\
\end{tabular}
\end{ruledtabular}
\end{table*}

\begin{figure}[!t]
    \centering
    \includegraphics[width=0.85\textwidth]{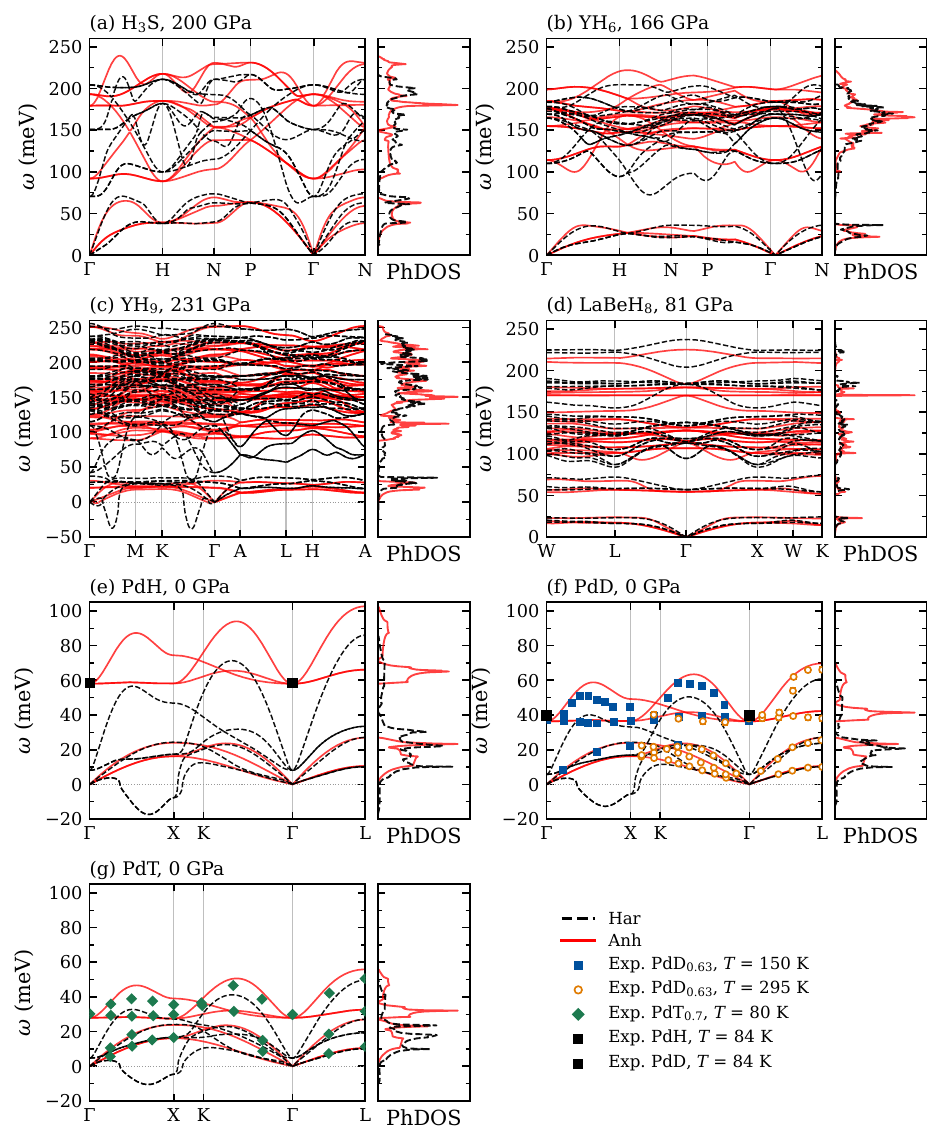}
    \caption{Harmonic DFPT (dashed) and anharmonic ML/SSCHA (solid) phonon dispersions and phonon DOS for (a) H$_3$S, (b) YH$_6$, (c) YH$_9$, (d) LaBeH$_8$, (e) PdH, (f) PdD, and (g) PdT. Symbols mark experimental Raman maxima for PdH and PdD~\cite{Sherman1977}, plotted at $\Gamma$ for reference, and inelastic neutron scattering measurements for nonstoichiometric PdD$_{0.63}$~\cite{Rowe1974} and PdT$_{0.70}$~\cite{Rowe1986}.} 
    \label{fig:phonons}
\end{figure}

Figure~\ref{fig:phonons} compares the harmonic DFPT and anharmonic ML/SSCHA phonon dispersions and densities of states. Anharmonicity produces an overall hardening of the phonon spectra in all systems, although the magnitude of the renormalization is strongly material dependent. The shifts are substantial throughout the H-derived optical branches of H$_3$S, YH$_6$, and YH$_9$, and the small imaginary pockets in the harmonic YH$_9$ spectrum are removed. LaBeH$_8$ exhibits the weakest renormalization, with comparatively modest changes in its phonon branches. In the Pd hydrides, anharmonicity stabilizes the lattice and hardens the H-derived rattling modes, with the largest shift occurring for the lightest isotope~\cite{Errea2013,Worsham1957,Bianco2026}. 
The PdH/D/T spectra agree well with the Raman and inelastic neutron scattering measurements~\cite{Sherman1977,Rowe1974,Rowe1986}. 

\begin{figure}[!t]
    \centering
    \includegraphics[width=0.6\textwidth]{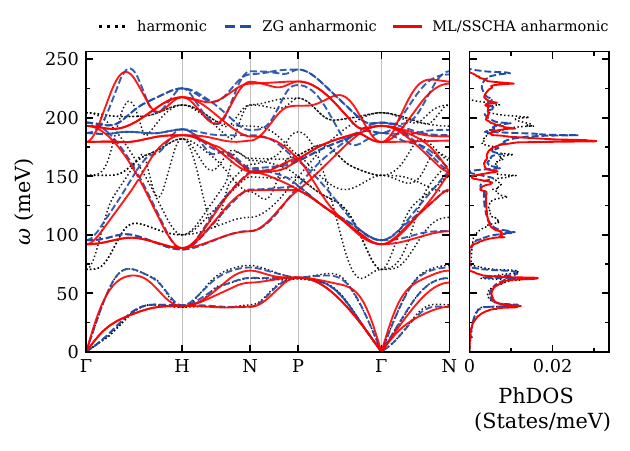}
    \caption{Anharmonic phonon dispersion and phonon DOS of H$_3$S at 200~GPa computed with the ZG special-displacement method (blue, dashed)~\cite{Mishra2025} and with the present ML/SSCHA workflow (red, solid), using the harmonic DFPT dispersion as reference (grey, dotted).}
    \label{fig:zgml}
\end{figure}

Figure~\ref{fig:zgml} benchmarks the present ML/SSCHA phonons of H$_3$S against the ZG special-displacement results used in our previous work~\cite{Mishra2025,Mishra2026}. The two approaches agree closely for the acoustic and low-energy optical branches and yield nearly identical hardening of the H--S stretching modes. The largest differences occur in the highest optical branches near the zone boundary, where the ZG frequencies are up to approximately 10~meV larger. This agreement supports the use of ML/SSCHA as a consistent anharmonic framework for all materials studied here.

\section{Electron-phonon and Superconducting properties}
\label{sec:supercond}

For interpreting adiabaticity trends, we define the effective electronic scale $\efeff$ entering $\eta=\wlg^{\rm anh}/\efeff$ as the half-width of the energy window around $\ef$ over which the DOS remains within $\pm15\%$ of $N(\ef)$. A small $\efeff$ therefore signals rapid DOS variation associated with features such as a van Hove singularity, band edge, or shallow pocket near $\ef$, which can increase the susceptibility to nonadiabatic effects. For H$_3$S, this criterion gives $\efeff\approx40$~meV, set by the van Hove singularity just below $\ef$~\cite{Mishra2026}, consistent with the 40--50~meV effective Fermi-energy scale inferred from Fermi-velocity arguments~\cite{Gorkov2018,Talantsev2022}. Table~\ref{tab:parameters} lists $\efeff$ for the hydrides considered here, together with the vertex ratio $\RV=|\lV_{\rm anh}|/\l_{\rm anh}$ and the adiabaticity ratio $\eta$.

\begin{table}[!hbt]
\caption{Electron-phonon parameters calculated in this work. Here, $\l_{\rm anh}$ and $\lV_{\rm anh}$ denote the linear and vertex corrected e-ph coupling for anharmonic phonons, $\RV=|\lV_{\rm anh}|/\l_{\rm anh}$ the corresponding vertex ratio, $\efeff$ the effective electronic energy scale, and $\eta=\wlg^{\rm anh}/\efeff$ the adiabaticity ratio.}
\label{tab:parameters}
\begin{ruledtabular}
\begin{tabular}{l c c c c c c}
Material & $\l_{\rm anh}$ & $\lV_{\rm anh}$ & $\RV$ & $\wlg^{\rm anh}$ (meV) & $\efeff$ (eV) & $\eta$ \\
\colrule
H$_3$S    & 1.49 & 0.48   & 0.32 & 127 & 0.04 & 3.5 \\
YH$_6$    & 1.74 & 0.25   & 0.14 & 116 & 0.63 & 0.18 \\
YH$_9$    & 1.83 & 0.26   & 0.14 & 112 & 0.45 & 0.25 \\
LaBeH$_8$ & 1.62 & 0.17   & 0.11 & 85  & 0.22 & 0.39 \\
PdH       & 0.48 & 0.040  & 0.08 & 44 & 0.41 & 0.11 \\
PdD       & 0.57 & 0.044  & 0.08 & 33 & 0.41 & 0.08 \\
PdT       & 0.62 & 0.052  & 0.08 & 27 & 0.41 & 0.07 \\
\end{tabular}
\end{ruledtabular}
\end{table}

\begin{figure}[!hbt]
    \centering
    \includegraphics[width=\textwidth]{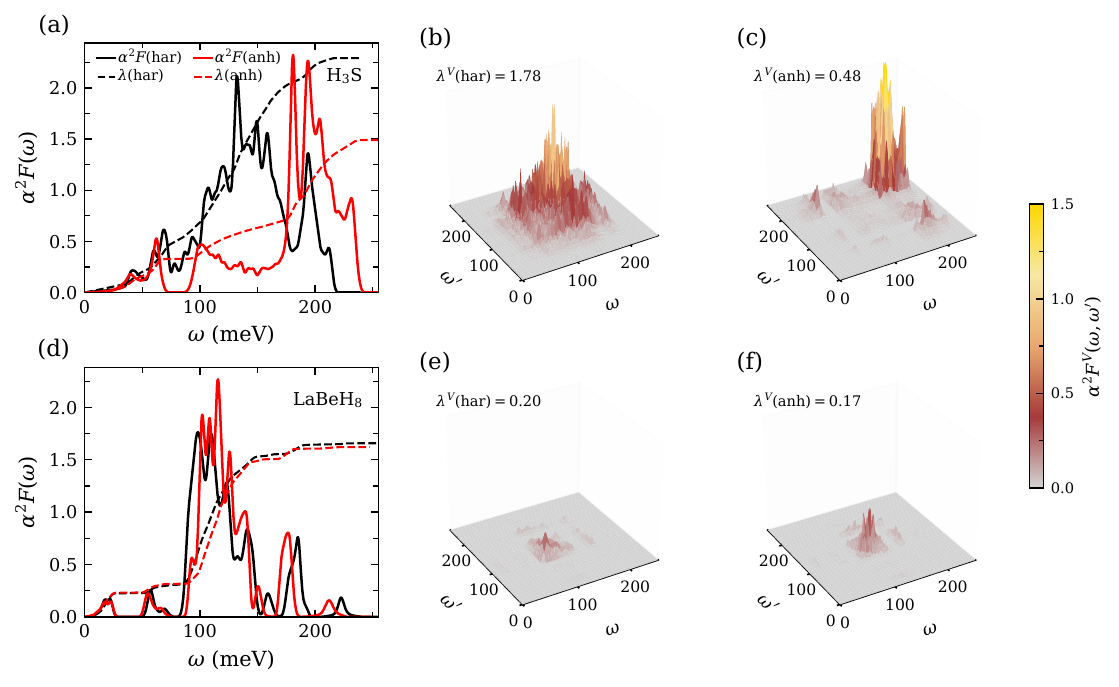}
    \caption{Eliashberg spectral function \aaff{} (solid), cumulative $\l(\o)$ (dashed), and vertex spectral functions \aaffv{} for (a,b,c) H$_3$S and (d,e,f) LaBeH$_8$, using harmonic and anharmonic phonons as indicated. }
    \label{fig:3D-vertex}
\end{figure}

\begin{figure}[!hbt]
    \centering
    \includegraphics[width=\textwidth]{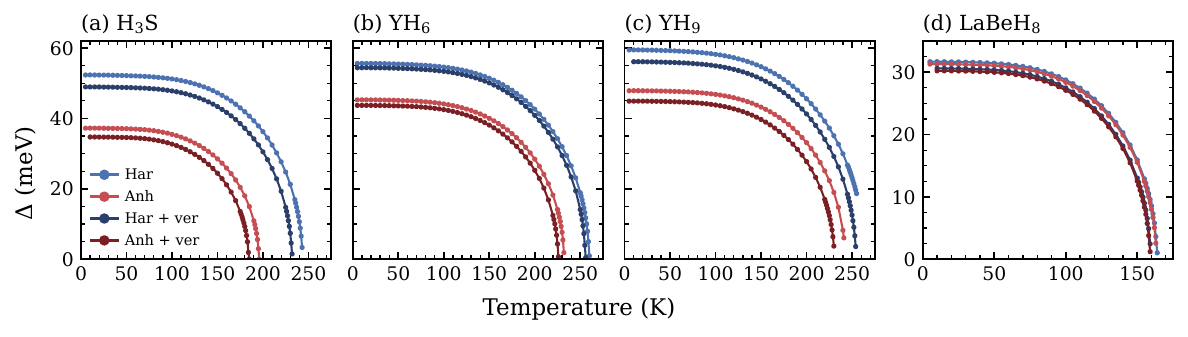}
    \caption{Isotropic superconducting gap functions $\Delta(T)$ calculated with the FBW approach for (a)~H$_3$S, (b)~YH$_6$, (c)~YH$_9$, and (d)~LaBeH$_8$. Results are shown for harmonic and anharmonic phonons, with and without first-order vertex corrections, using $\mu^{*}=0.13$.}
    \label{fig:gaps}
\end{figure}

\begin{figure}[!hbt]
    \centering
    \includegraphics[width=\textwidth]{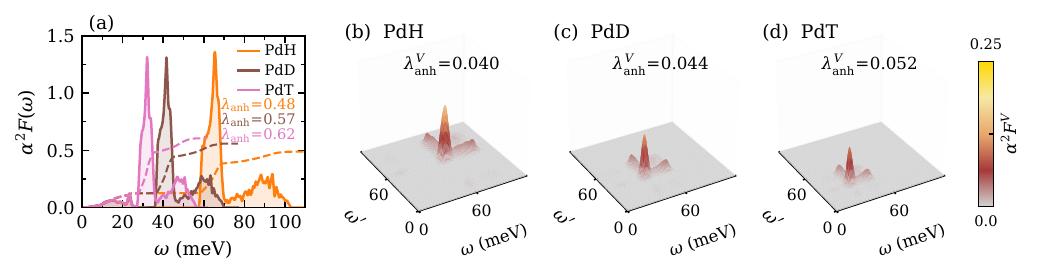}
    \caption{Anharmonic (a) $\a^2F(\o)$ and (b--d) $\a^2F^{V}(\o,\o')$ for PdH, PdD, and PdT. The integrated vertex couplings are $\lV=0.040$, 0.044, and 0.052, giving $\RV\approx0.08$. }
    \label{fig:pd-vertex}
\end{figure}

\begin{figure}[!hbt]
    \centering
    \includegraphics[width=\textwidth]{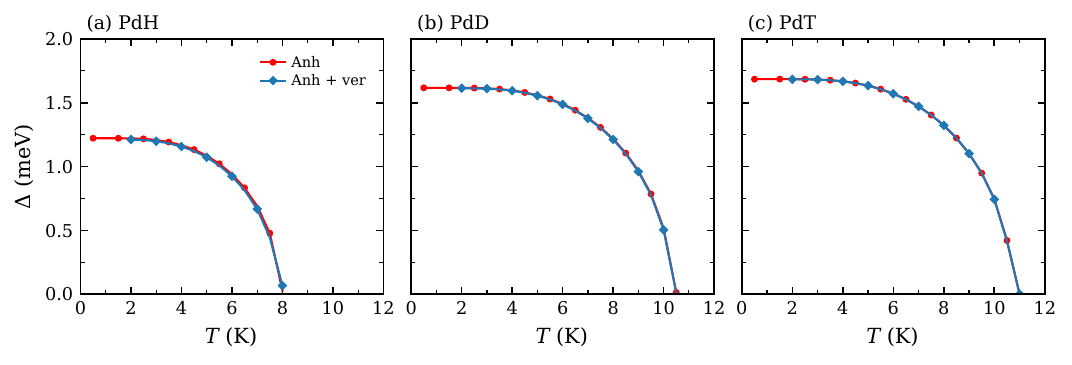}
    \caption{Isotropic superconducting gap functions $\Delta(T)$ calculated with the FBW approach for (a)~PdH, (b)~PdD, and (c)~PdT. Results are shown for anharmonic phonons, with and without first-order vertex corrections, using $\mu^{*}=0.085$.}
    \label{fig:pdgap}
\end{figure}

Figure~\ref{fig:3D-vertex} shows the effect of anharmonicity on the spectral functions and e-ph couplings of H$_3$S and LaBeH$_8$. In H$_3$S, anharmonicity reshapes \aaff{}, concentrating spectral weight in a pronounced high-frequency peak while reducing the cumulative coupling from $\l_{\rm har}=2.29$ to $\l_{\rm anh}=1.49$ [Fig.~\ref{fig:3D-vertex}(a)]. The harmonic vertex spectral function is broadly distributed over the two-frequency plane, giving $\lV_{\rm har}=1.78$ [Fig.~\ref{fig:3D-vertex}(b)]. Anharmonicity localizes the spectral weight in a narrower high-frequency region and suppresses the broad background, reducing the integrated vertex coupling to $\lV_{\rm anh}=0.48$ [Fig.~\ref{fig:3D-vertex}(c)]. In LaBeH$_8$, anharmonicity only modestly reshapes the conventional \aaff{}, reducing the coupling from $\l_{\rm har}=1.66$ to $\l_{\rm anh}=1.62$ [Fig.~\ref{fig:3D-vertex}(d)], consistent with the modest renormalization found in the ML-potential SSCHA study of Ref.~\cite{Dong2025}. The harmonic vertex spectral function is confined largely to the H-derived optical-frequency range and gives $\lV_{\rm har}=0.20$ [Fig.~\ref{fig:3D-vertex}(e)]. Anharmonicity redistributes this spectral weight but changes its integral only slightly, lowering the vertex coupling to $\lV_{\rm anh}=0.17$ [Fig.~\ref{fig:3D-vertex}(f)]. The resulting vertex correction remains small, with $\RV=0.11$.

Figure~\ref{fig:gaps} shows the isotropic superconducting gaps $\Delta(T)$ for the high-pressure hydrides using harmonic and anharmonic phonons, with and without first-order vertex corrections. Anharmonicity substantially reduces both the low-temperature gap and $\tc$ in H$_3$S, YH$_6$, and YH$_9$, while the vertex correction provides an additional suppression. The separation among the four treatments is largest in H$_3$S, where the vertex correction lowers $\tc$ by 12~K, comparable to the reduction found in our earlier work~\cite{Mishra2025,Mishra2026}. The separation is also pronounced in YH$_9$ but considerably smaller in YH$_6$. In LaBeH$_8$, anharmonicity has a negligible impact, lowering $\tc$ by only 1~K, consistent with its weak anharmonic phonon renormalization. The first-order vertex contribution is somewhat larger but remains modest, reducing the anharmonic FBW $\tc$ by 4~K.

\begin{figure}[!hbt]
    \centering
    \includegraphics[width=0.45\textwidth]{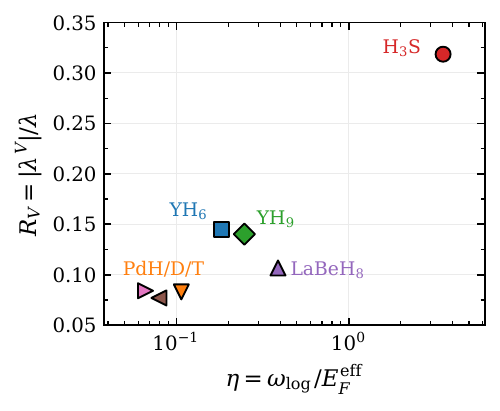}
    \caption{Vertex ratio $\RV=|\lV_{\rm anh}|/\l_{\rm anh}$ versus the adiabaticity parameter $\eta=\omega_{\rm log}^{\rm anh}/\efeff$. All values are tabulated in Table~\ref{tab:parameters}.}
    \label{fig:eta-rv}
\end{figure}

Figure~\ref{fig:pd-vertex} shows the anharmonic spectral functions for the ambient-pressure Pd hydrides. The conventional \aaff{} is dominated by H-derived optical modes that shift to lower frequencies with increasing isotope mass, yielding $\l_{\rm anh}=0.48$, 0.57, and 0.62 for PdH, PdD, and PdT, respectively. The vertex spectra are confined primarily to the same frequency range, with $\lV_{\rm anh}=0.040$, 0.044, and 0.052, respectively. Relative to the corresponding conventional couplings, these values give $\RV\approx0.08$ for each isotope, the smallest values among the materials studied. Figure~\ref{fig:pdgap} compares the FBW gap functions with and without the first-order vertex correction. The curves nearly overlap, showing only a slight suppression of $\Delta(T)$ and changes in $\tc$ below 0.05~K.

Figure~\ref{fig:eta-rv} compares the vertex ratio $\RV=|\lV_{\rm anh}|/\l_{\rm anh}$ with the adiabaticity ratio $\eta=\omega_{\rm log}^{\rm anh}/\efeff$. The ratio $\eta$ compares the characteristic phonon and electronic energy scales and thus serves as an energy-scale diagnostic of nonadiabaticity, whereas $\RV$ directly quantifies the calculated vertex contribution obtained from the full two-frequency spectral function. H$_3$S, whose van Hove-limited $\efeff\approx40$~meV places it at $\eta\approx3.5$, is the only material in the nonadiabatic regime $\eta\gtrsim1$ and has the largest vertex ratio, $\RV=0.32$. At the opposite extreme, the Pd hydrides have the smallest values of both ratios, with $\eta\approx0.08$ and $\RV\approx0.08$, placing them closest to the adiabatic limit. As discussed above, the first-order vertex term weakens the pairing kernel in all systems studied here and either lowers $\tc$ or leaves it essentially unchanged.

\begin{figure}[!hbt]
    \centering
    \includegraphics[width=0.95\textwidth]{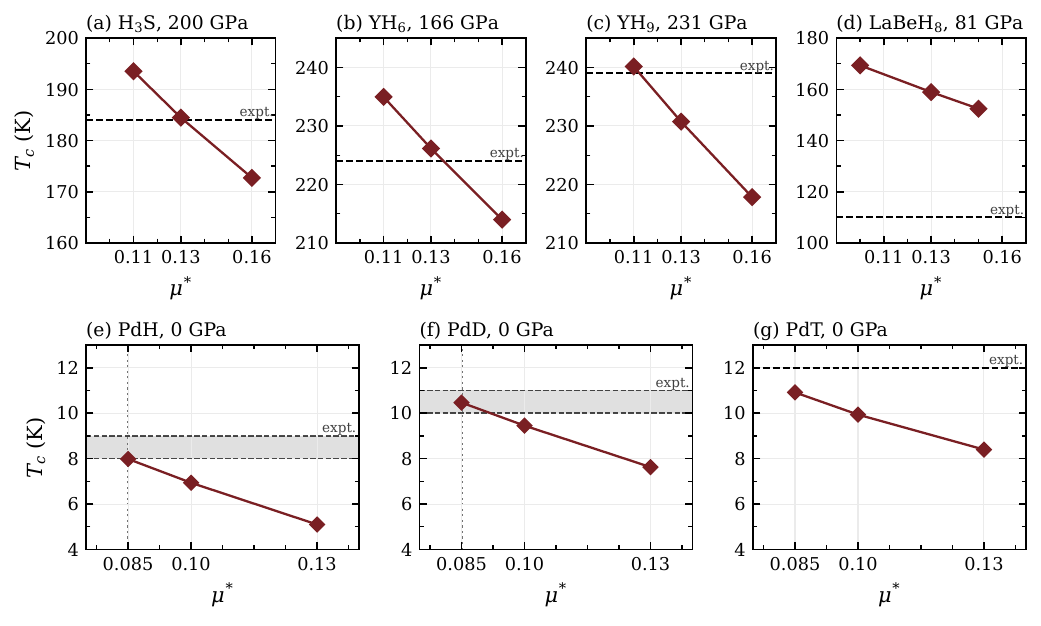}
    \caption{Dependence of $\tc$ on the Coulomb pseudopotential $\mu^{*}$ calculated within the FBW formalism using anharmonic phonons and first-order vertex corrections for (a)~H$_3$S, (b)~YH$_6$, (c)~YH$_9$, (d)~LaBeH$_8$, (e)~PdH, (f)~PdD, and (g)~PdT. The dashed horizontal lines and shaded regions indicate the experimental $\tc$ values and ranges reported for H$_3$S at 200~GPa~\cite{Drozdov2015}, YH$_6$ at 166~GPa~\cite{Troyan2021}, YH$_9$ at 231~GPa~\cite{Kong2021}, LaBeH$_8$ at 80~GPa~\cite{Song2023}, and PdH, PdD, and PdT at 0~GPa~\cite{Skoskiewicz1972,Stritzker1972,Rowe1986}. }
    \label{fig:muc}
\end{figure}

Figure~\ref{fig:muc} illustrates the sensitivity of the $\tc$ values to the Coulomb pseudopotential $\mu^{*}$, as calculated within the FBW formalism using anharmonic phonons and first-order vertex corrections. For H$_3$S, YH$_6$, and YH$_9$, the calculated values remain close to experiment near the conventional choice $\mu^{*}=0.13$. By contrast, LaBeH$_8$ remains substantially above experiment over the entire range shown; a linear extrapolation of the computed values indicates that reproducing the measured $\tc$ of 110~K would require $\mu^{*}\approx0.28$, far outside the conventional range.
For PdH, PdD, and PdT, we retain $\mu^{*}=0.085$ from the anharmonic SSCHA study of Ref.~\cite{Errea2013}. Increasing $\mu^{*}$ to 0.13 would place the calculated $\tc$ values approximately 40\% below experiment for PdH and 25--30\% below experiment for PdD and PdT.

Table~\ref{tab:literature} presents a comparison of our results with previous calculations and experiments. For H$_3$S at 200~GPa, Errea \textit{et al.}~\cite{Errea2015} obtained $\l_{\rm anh}=1.84$, $\wlg^{\rm anh}=92.9$~meV, and $\tc=194$~K using SSCHA phonons with $\mu^{*}=0.16$. Our earlier calculations using ZG special-displacement phonons yielded $\l_{\rm anh}=1.72$, $\wlg^{\rm anh}=110$~meV, and $\tc=189$~K without and 178~K with the vertex correction at the same $\mu^{*}=0.16$~\cite{Mishra2025,Mishra2026}. The present ML/SSCHA treatment gives a smaller coupling, $\l_{\rm anh}=1.49$, but a higher $\wlg^{\rm anh}=127$~meV, yielding $\tc=196$~K without and 184~K with the vertex correction at $\mu^{*}=0.13$. Thus, all anharmonic calculations without vertex corrections remain above experiment, whereas including vertex effects lowers $\tc$. The earlier ZG-based result lies slightly below the measured value, while the present calculation reproduces the experimental $\tc=184$~K at 200~GPa~\cite{Drozdov2015}.

For YH$_6$ at 165~GPa, Troyan \textit{et al.}~\cite{Troyan2021} obtained harmonic and anharmonic $\tc$ values of 261--272 and 236--247~K, respectively, using SSCHA phonons with $\mu^{*}=0.10$--0.15, and reported $\l_{\rm anh}=1.71$. Ding \textit{et al.}~\cite{Ding2025} obtained the same e-ph coupling with anharmonic phonons using the stochastic path-integral approach (SPIA) but predicted a lower $\tc=212$~K with $\mu^{*}=0.11$. Our anharmonic FBW calculation at the experimental pressure of 166~GPa gives $\l_{\rm anh}=1.74$ and $\tc=233$~K with $\mu^{*}=0.13$. Including the first-order vertex correction further lowers $\tc$ to 226~K, close to the measured value of 224~K at 166~GPa~\cite{Troyan2021}. Harmonic FBW results at 200~GPa obtained with $\mu^{*}=0.16$~\cite{Lucrezi2024} and with \emph{ab initio} Coulomb treatments~\cite{Kogler2025} likewise exceeded the experimental value.

For YH$_9$ at 231~GPa, our anharmonic calculation gives $\lambda_{\rm anh}=1.83$ and $\tc=244$~K for $\mu^{*}=0.13$, with the first-order vertex correction reducing $\tc$ to 231~K, compared with the measured 239~K. Using $\mu^{*}=0.11$ instead raises the vertex-corrected value to 240~K, essentially reproducing the experimental value [Fig.~\ref{fig:muc}]~\cite{Kong2021}. Ding \textit{et al.}~\cite{Ding2025} obtained $\lambda_{\rm anh}=1.75$ and $\tc=240$~K with $\mu^{*}=0.11$, but at a higher pressure of 255~GPa.

LaBeH$_8$ remains the notable exception to the overall agreement with experiment, as its calculated $\tc$ remains substantially too high even after including anharmonic and first-order vertex corrections. The vertex correction lowers the $\tc$ obtained with anharmonic phonons by only 4~K, from 163 to 159~K, compared with the measured value of approximately 110~K, and the discrepancy persists throughout the $\mu^{*}$ range shown in Fig.~\ref{fig:muc}. Previous calculations have likewise remained above experiment, reporting harmonic and anharmonic Eliashberg estimates of 131--167~K over the range $\mu^{*}=0.10$--0.15~\cite{Wang2024,Dong2025,Kogler2025}.

The comparison is more involved for the Pd hydrides. Earlier anharmonic SSCHA calculations yielded $\tc=5.0$, 6.5, and 6.9~K for PdH, PdD, and PdT, respectively, using the same $\mu^{*}=0.085$~\cite{Errea2013}. Their couplings, $\l_{\rm anh}=0.40$, 0.46, and 0.48, are lower than our values of 0.48, 0.57, and 0.62. At the same time, their $\omega_{\log}^{\rm anh}$ values of 50.2, 37.7, and 31.9~meV are higher than our corresponding values of 44.0, 33.0, and 27.0~meV. Comparison of the anharmonic phonon dispersions shows broadly similar acoustic branches, with the principal differences occurring in the H-derived optical modes. In particular, the lowest optical branch is softer in our calculations and appears closer to the available measurements for PdD$_{0.63}$ and PdT$_{0.70}$~\cite{Rowe1974,Rowe1986} as shown in Fig.~\ref{fig:phonons}. Because spectral weight at lower frequencies contributes more strongly to $\l$, this feature accounts for much of the larger coupling. The increase in $\l_{\rm anh}$ offsets the lower $\omega_{\log}^{\rm anh}$ and yields higher $\tc$ values. Finally, a recent study reported linear e-ph couplings of 0.42 and 0.48 for PdH and PdD, respectively, close to the earlier values~\cite{Bianco2026}. Despite appreciable differences in the anharmonic optical branches, the similar linear couplings and $\omega_{\log}$ values indicate that these spectral differences largely average out in the integrated e-ph quantities. Including nonperturbative nonlinear e-ph vertices increased the total couplings to 0.64 and 0.72 and yielded Allen-Dynes estimates of $\tc=11.0$ and 13.0~K for PdH and PdD, respectively, using $\mu^{*}=0.13$~\cite{Bianco2026}.

\begin{table}[!hbt]
\caption{Representative first-principles results for the hydrides studied here. Values are quoted at the indicated pressures and Coulomb pseudopotentials $\mu^{*}$. For entries containing paired PdH/PdD values, the first and second values correspond to PdH and PdD, respectively. Abbreviations: SSCHA = stochastic self-consistent harmonic approximation; ML/SSCHA = machine-learning accelerated SSCHA; ZG = Zacharias-Giustino special-displacement method; SPIA = stochastic path-integral approach; MTP = moment tensor potential; aniso = anisotropic Migdal-Eliashberg; ``$+$ vertex'' = including the first-order vertex correction. 
For Ref.~\cite{Kogler2025}, $\mu^{*}$ is obtained from the \textit{ab initio} Coulomb parameter $\mu$, computed from the $GW$ screened interaction, using the Morel-Anderson renormalization $\mu^{*}=\mu/[1+\mu\ln(\varepsilon_{\rm el}/\omega_{\rm c})]$, where $\omega_{\rm c}$ is the Matsubara frequency cutoff and $\varepsilon_{\rm el}$ the electronic bandwidth. $\dagger$ marks $\tc$ obtained directly with the bare $\mu$, and $W$ marks $\tc$ obtained with the full energy-dependent screened Coulomb interaction $W(\epsilon,\epsilon')$.
Calculated results are quoted at static-lattice pressures and experimental results at the measured pressures.} 
\label{tab:literature}
\footnotesize
\setlength{\tabcolsep}{4pt}
\begin{ruledtabular}
\begin{tabular}{l l c l c c c c}
System & Study & $P$ (GPa) & Phonons & $\l$ & $\wlg$ (meV) & $\mu^{*}$ & $\tc$ (K) \\
\colrule
H$_3$S
& Errea 2015~\cite{Errea2015}
& 200 & SSCHA & 1.84 & 92.9 & 0.16 & 194 \\
& Mishra 2025/2026~\cite{Mishra2025,Mishra2026}
& 200 & ZG & 1.72 & 110 & 0.16 & 189 \\
& & & ZG $+$ vertex & 1.72 & 110 & 0.16 & 178 \\
& This work
& 200 & harmonic & 2.29 & 107 & 0.13 & 244 \\
& & & harmonic $+$ vertex & 2.29 & 107 & 0.13 & 232 \\
& & & ML/SSCHA & 1.49 & 127 & 0.13 & 196 \\
&  &  & ML/SSCHA $+$ vertex & 1.49 & 127 & 0.13 & 184 \\
& Experiment~\cite{Drozdov2015} & 200 & -- & -- & -- & -- & 184 \\
\colrule
YH$_6$
& Troyan 2021~\cite{Troyan2021}
& 165 & harmonic & 2.24 & 80 & 0.10--0.15 & 261--272 \\
& & & SSCHA & 1.71 & 114.9 & 0.10--0.15 & 236--247 \\
& Lucrezi 2024~\cite{Lucrezi2024}
& 200 & harmonic (aniso) & 2.0 & 108 & 0.16 & 238 \\
& Kogler 2025~\cite{Kogler2025}
& 200 & harmonic & -- & -- & 0.18 & 231 \\
& & & & -- & -- & $0.20^\dagger$ & 226 \\
& & & & -- & -- & $W$ & 227 \\
& Ding 2025~\cite{Ding2025}
& 165 & SPIA & 1.71 & -- & 0.11 & 212 \\
& This work
& 166 & harmonic & 2.36 & 101 & 0.13 & 260 \\
& & & harmonic $+$ vertex & 2.36 & 101 & 0.13 & 256 \\
& & & ML/SSCHA & 1.74 & 116 & 0.13 & 233 \\
& &  & ML/SSCHA $+$ vertex & 1.74 & 116 & 0.13 & 226 \\
& Experiment~\cite{Troyan2021} & 166 & -- & -- & -- & -- & 224 \\
\colrule
YH$_9$
& Ding 2025~\cite{Ding2025}
& 255 & SPIA & 1.75 & -- & 0.11 & 240 \\
& This work
& $231$ & harmonic & 3.09 & 72 & 0.11 & 274 \\
& & & harmonic $+$ vertex & 3.09 & 72 & 0.11 & 263 \\
& & & ML/SSCHA  & 1.83 & 112 & 0.11 & 252 \\
& & & ML/SSCHA $+$ vertex & 1.83 & 112 & 0.11 & 240 \\
& & & harmonic & 3.09 & 72 & 0.13 & 265 \\
& & & harmonic $+$ vertex & 3.09 & 72 & 0.13 & 255 \\
& & & ML/SSCHA  & 1.83 & 112 & 0.13 & 244 \\
& & & ML/SSCHA $+$ vertex & 1.83 & 112 & 0.13 & 231 \\
& Experiment~\cite{Kong2021} & 231 & -- & -- & -- & -- & 239 \\
\colrule
LaBeH$_8$
& Wang 2024~\cite{Wang2024}
& 80 & harmonic & 1.60 & 88.5 & 0.10/0.13 & 167/154 \\
& Kogler 2025~\cite{Kogler2025}
& 100 & harmonic & -- & -- & 0.15 & 141 \\
& & & & -- & -- & $0.16^*$ & 140 \\
& & & & -- & -- & $\mu^\dagger$ & 140 \\
& Dong 2025~\cite{Dong2025}
& 100 & SSCHA $+$ MTP & 1.59 & 90.1 & 0.10--0.15 & 138--160 \\
& This work
& 81 & harmonic & 1.66 & 85 & 0.13 & 164 \\
& & & harmonic $+$ vertex & 1.66 & 85 & 0.13 & 159 \\
& & & ML/SSCHA & 1.62 & 85 & 0.13 & 163 \\
& & & ML/SSCHA $+$ vertex & 1.62 & 85 & 0.13 & 159 \\
& Experiment~\cite{Song2023} & 80 & -- & -- & -- & -- & 110 \\
&  & 100 & -- & -- & -- & -- & 105 \\
\colrule
PdH/PdD
& Errea 2013~\cite{Errea2013}
& 0 & SSCHA & 0.40/0.46 & -- & 0.085 & 5.0/6.5 \\
& Bianco 2026~\cite{Bianco2026}
& 0 & SSCHA, avg.\ nonlinear vertices & 0.64/0.72 & -- & 0.13 & 11/13 \\
& This work
& 0 & ML/SSCHA & 0.48/0.57 & 44/33 & 0.085 & 8.0/10.5 \\
& & & ML/SSCHA $+$ vertex  & 0.48/0.57 & 44/33 & 0.085 & 8.0/10.5 \\
& Experiment~\cite{Skoskiewicz1972,Stritzker1972,Rowe1986} & 0 & -- & -- & -- & -- & 8--9/10--11 
\end{tabular}
\end{ruledtabular}
\end{table}
\newpage
\bibliography{pap}